\RequirePackage{fix-cm}
\documentclass[smallextended,numbook]{svjour3a}       
\smartqed  
\usepackage{hyperref}
\usepackage[]{natbib}
\usepackage{url}
\makeatletter
\def\url@leostyle{%
	\def\UrlFont{\sf}}{\def\UrlFont{\small\ttfamily}}
\makeatother
\usepackage{amsmath}
\usepackage{amsfonts}
\usepackage{amssymb}

\usepackage[left=3cm,right=2cm,top=2.5cm,bottom=2cm,includeheadfoot]{geometry}

\usepackage[capitalise]{cleveref}

\usepackage{seqsplit}
\usepackage{tikz-cd}

\defcitealias{Abbott2016}{Abbott et. al., (2016)}

\newcommand{\Tr}{\operatorname{Tr}}%
\newcommand{\ZT}[1]{\textquotedblleft#1\textquotedblright}%

\newcommand{\dif}{\mathrm{d}}%
\newcommand{\ii}{\mathrm{i}}%

\newlength{\myl}%
\newcommand{\SUM}[2]{{\setlength{\myl}{\widthof{$\displaystyle\sum_{#1}^{#2}$}*\real{0.5}-\widthof{$\displaystyle\sum$}*\real{0.5}}\sum_{#1}^{#2}\;\hspace{-\the\myl}}}
\newcommand{\INT}[3]{\settowidth{\myl}{$\displaystyle\int_{#1}^{#2}$}{\int_{#1}^{#2}\;\;\;\hspace{-\the\myl}\dif #3}\,}
\newcommand{\TINT}[3]{\settowidth{\myl}{$\int_{#1}^{#2}$}{\int_{#1}^{#2}\!\ifthenelse{\equal{#1#2}{}}{}{\;\;\;\;\hspace{-\the\myl}}\dif #3}\,}%
\newcommand{\EINT}[3]{\settowidth{\myl}{$\int_{#1}^{#2}$}{\int_{#1}^{#2}\;\;\;\,\hspace{-\the\myl}\dif #3}\,}
\newcommand{\fdif}{\operatorname{\delta}}%
\newcommand{\Fdif}[2]{\frac{\fdif\!#1}{\fdif\!#2}}%
\newcommand{\Nabla}{\vec{\nabla}}%
\usepackage{graphicx}
\usepackage{braket}
\usepackage{mathabx}
\usepackage{color}
\begin{document}
	
	\title{Is thermodynamics fundamental?}
	
	\titlerunning{Is thermodynamics fundamental?}        
	
	\author{Michael te Vrugt$^{1,\star}$, Paul Needham$^2$, Georg J. Schmitz$^3$}
	
	\authorrunning{Michael te Vrugt, Paul Needham, and Georg J. Schmitz} 
	
	\institute{$^1$Institut f\"ur Theoretische Physik, Center for Soft Nanoscience, Philosophisches Seminar, Westf\"alische Wilhelms-Universit\"at M\"unster, D-48149 M\"unster, Germany\\
		$^2$Department of Philosophy, University of Stockholm, SE-106 91 Stockholm, Sweden\\
		$^3$MICRESS group, ACCESS e.V., Intzestr. 5, D-52072 Aachen, Germany\\
		$^\star$\email{michael.tevrugt@uni-muenster.de}           
	}
	
	\date{}
	
	\maketitle
	
	\begin{abstract}
		It is a common view in philosophy of physics that thermodynamics is a non-fundamental theory. This is motivated in particular by the fact that thermodynamics is considered to be a paradigmatic example for a theory that can be reduced to another one, namely statistical mechanics. For instance, the statement \ZT{temperature is mean molecular kinetic energy} has become a textbook example for a successful reduction, despite the fact that this statement is not correct for a large variety of systems. In this article, we defend the view that thermodynamics is a fundamental theory, a position that we justify based on four case studies from recent physical research. We explain how entropic gravity (1) and black hole thermodynamics (2) can serve as case studies for the multiple realizability problem which blocks the reduction of thermodynamics. Moreover, we discuss the problem of the reducibility of phase transitions and argue that bifurcation theory (3) allows the modelling of \ZT{phase transitions} on a thermodynamic level even in finite systems. It is also shown that the derivation of irreversible transport equations in the Mori-Zwanzig formalism (4) does not, despite recent claims to the contrary, constitute a reduction of thermodynamics to quantum mechanics. Finally, we briefly discuss some arguments against the fundamentality of thermodynamics that are not based on reduction. 
	\end{abstract}
	\section{Introduction}
	It is beyond doubt that thermodynamics is among the most important physical theories. This is evident both from the prominent role it plays in the physics curriculum and in physical research and from its numerous applications in chemistry and engineering. But whether it is a fundamental theory is a much more subtle question. Prima facie, there appears to be plenty of evidence for this view. Many physicists appear to be more confident in the validity of the second law of thermodynamics than in anything else their discipline has ever achieved. Thermodynamics is used to learn about black holes and quantum gravity, areas where little to no microscopic knowledge is currently available---implying that we are confident that it still holds there. And finally, philosophers of physics (who typically show much more interest in fundamental theories like general relativity or quantum mechanics than in non-fundamental ones) have spent a considerable amount of time thinking about thermodynamics.
	
	On the other hand, many philosophers of physics have come to the conclusion that thermodynamics is \textit{not} a fundamental theory. This view has two closely connected motivations. First, is argued that thermodynamics can be reduced to quantum (or classical) mechanics in the framework of statistical mechanics \citep{Nagel1961}. Second, it is argued that thermodynamics is only approximately correct, for example due to the fact that its statements about phase transitions are true only for infinite systems \citep{Callender2001}. The reduction of thermodynamics is, however, not an easy task. A particularly important problem in this context is \textit{multiple realizability}. This problem, originally studied in the philosophy of mind \citep{Putnam1967}, has also received some attention in the philosophy of thermodynamics \citep{Batterman2000}. For example, it is frequently claimed that \ZT{temperature} (a macroscopic property of thermodynamic systems) could be reduced to \ZT{mean molecular kinetic energy} (a microscopic property). However, this microscopic definition of temperature holds only for gases, whereas numerous other systems realize \ZT{temperature} in very different ways \citep{Bickle1998,Needham2010}.
	
	The discussions of this problem and its implications for the reducibility of thermodynamics typically concern a rather basic level of physics, being concerned with temperature and occasionally with phase transitions. This stands in marked contrast to other sub-fields of the philosophy of thermal physics, such as black hole thermodynamics \citep{Wallace2018,Wallace2019,Curiel2014}, the study of thermodynamic irreversibility \citep{Wallace2015,Robertson2018,teVrugt2021c}, or even the metaphysics of statistical mechanics \citep{teVrugt2021b}, which often take place on a considerably technical level and take into account very recent results from research in physics. To close this gap, the present work discusses the fundamentality of thermodynamics using four case studies, namely entropic gravity (1), black hole thermodynamics (2), phase transitions in spatially resolved thermodynamics (3), and the derivation of irreversible master equations in the Mori-Zwanzig formalism (4). Based on these diverse physical results, we argue that thermodynamics is a fundamental theory that cannot be reduced to statistical mechanics. The results of our discussion are, as will be discussed below, relevant also for more general debates in philosophy of science and physics, such as the multiple realizability problem, the problem of the thermodynamic limit and the problem of the arrow of time. For instance, it is shown how modern developments of thermodynamics \citep{ThieleFHEKA2019} facilitate the description of phase transitions without the infinite idealization usually thought to be required here.
	
	This article is structured as follows. In \cref{fot}, we introduce the problem of the fundamentality of thermodynamics. \cref{uff} explains how \ZT{fundamentality} is understood in this work. In \cref{ramr}, we address the multiple realizability problem. Case studies in physics are then provided in \cref{entropicgravity} (entropic gravity), \cref{cbht} (black hole thermodynamics), \cref{pt} (phase transitions), and \cref{arrow} (arrow of time). Some arguments against the fundamentality of thermodynamics are discussed in \cref{other}. The discussion is concluded in \cref{conclusion}.
	
	
	
	
	
	
	
	
	\section{\label{fot}Fundamentality of thermodynamics: Why bother?}
	Many physicists have considered thermodynamics to be one of the most fundamental theories in physics---if not \textit{the} most fundamental one. This view was expressed in a famous quote by \citet[p. 81]{Eddington1935}:
	
	\begin{quotation}
		The law that entropy always increases---the second law of thermodynamics---holds, I think, the supreme position among the laws of Nature. If someone points out to you that your pet theory of the universe is in disagreement with Maxwell’s equations---then so much the worse for Maxwell’s equations. If it is found to be contradicted by observation---well, these experimentalists bungle things sometimes. But if your theory is found to be against the second law of thermodynamics I can give you no hope; there is nothing for it but to collapse in deepest humiliation
	\end{quotation}
	
	A similar perspective has been taken by \citet[p. 33]{Einstein1970}\footnote{The quotes of Einstein and Eddington are cited from cited from \citet[p. 540]{Callender2001}.}
	\begin{quotation}
		[Classical thermodynamics] is the only theory of universal content concerning which I am convinced that, within the framework of the applicability of its basic concepts, it will never be overthrown.
	\end{quotation}
	
	Such a view can be found also in more application-focused treatments of thermodynamics:
	
	\begin{quotation}
		Together, continuum mechanics and thermodynamics form the fundamental theory at the heart of many disciplines in science and engineering \citep[p. xi]{TadmorME2012}
	\end{quotation}
	
	On the other hand, it is rather common---in particular in the philosophical foundations of thermodynamics---to see thermodynamics not as a fundamental theory. This is expressed in the following two quotes from the philosophical literature:
	
	\begin{quotation}
		But we don’t think that thermodynamics is a fundamental theory \citep[p. 312]{North2011}
	\end{quotation}
	
	\begin{quotation}
		But some people, it seems to me, have \textit{too much}
		respect for thermodynamics. They take the field, from a foundational perspective, \textit{too seriously}. And this taking of the field too seriously is responsible for significant errors in the foundations of statistical mechanics. \citep[p. 540]{Callender2001}
	\end{quotation}
	
	The latter position is typically based on two sorts of arguments. First, it is argued that thermodynamics holds only approximately or that there are exceptions. Second, it is argued that thermodynamics can be reduced to (classical or quantum) statistical mechanics. These two arguments are closely related since it is the reduction that is supposed to uncover the required corrections to thermodynamics.
	
	Hence, the fundamentality of thermodynamics appears to be an open and quite controversial topic. This problem is of possible importance also for philosophy in general. There is a long (and increasingly popular) tradition of using scientific theories as a basis for solving metaphysical problems, a line of research that is known as \ZT{inductive metaphysics} \citep{Scholz2019} or \ZT{scientific metaphysics} \citep{LadymanR2007}. Thermodynamics is no exception here and has inspired, e.g., work on the special composition question \citep{teVrugt2021} and the nature of processes \citep{Needham2013,Needham2018}. An argument that can be raised against such approaches is based on the premises that (a) thermodynamics might not be fundamental and (b) only fundamental sciences should be used in metaphysical research. While there are good reasons of rejecing (b) (see \citet{Needham2013,teVrugt2021,teVrugt2021b} for a discussion), the case for \ZT{thermodynamic metaphysics} can be made even stronger if premise (a) is also refuted.
	
	Another reason why this problem is important has to do with the status of thermodynamics and Einstein's view of it. Thermodynamics is a \ZT{principle theory} or \ZT{phenomenological theory}, i.e., it is not based on some microscopic model of matter, but simply rests on empirically very well-supported principles that are thought to hold for any system irrespective of its microscopic constitution. This means, in particular, that we can apply them even if we do not know the microscopic structure of the system we want to apply it to. As discussed in detail by \citet{Brown2005}, Einstein was inspired by this feature of thermodynamics when he invented special relativity. Around 1905, there had been a variety of attempts to explain, e.g., the null result of the Michelson-Morley experiment, based on considerations about the microscopic structure of the experimental system (which, however, was not known due to the status of condensed matter physics at that time). Instead of developing a microscopic theory, Einstein simply introduced two phenomenological postulates (the principle of relativity and the light postulate) that he took to hold universally, and thereby invented special relativity. Nowadays, it is still generally believed among physicists that whatever our microscopic theories look like, they have to obey the universal laws of relativity (although some philosophers, such as Brown himself, argue instead that we should see special relativity as emerging from microscopic dynamical laws). Thus, given that they have been introduced in quite a similar spirit, it seems a little strange that the theory of relativity is generally considered a fundamental theory whereas thermodynamics is not.
	
	It should be noted that in this paper we adopt a rather generous view of what counts as thermodynamics. In particular, we take it to include the minus first law of thermodynamics proposed by \citet{BrownU2001} (based on \citet{UhlenbeckF1963}), which captures the tendency of nonequilibrium systems to approach equilibrium, as well as various results that are discussed in physics as \textit{nonequilibrium thermodynamics} \citep{deGrootM1962}, \textit{stochastic thermodynamics} \citep{Seifert2012}, or \textit{quantum thermodynamics} \citep{VinjanampathyA2016}. Some authors (such as \citet{Myrvold2020b}) have argued for a more conservative understanding. We believe that our general concept is a more accurate reflection of how thermodynamics is understood in physics, but if you have reservations here, feel free to replace the word \ZT{thermodynamics} by something like \ZT{extended thermodynamics} every time it appears in the article, and to read it as arguing for the fundamentality of this extended theory. 
	
	\section{\label{uff}Understanding fundamentality}
	An important observation to make regarding the quotes on the fundamentality of thermodynamics from \cref{fot} is that they are not all based on the same idea of \ZT{fundamentality}. \citet{Eddington1935} and \citet{Einstein1970} for example (in whose quotes the word \ZT{fundamentality} does not appear explicitly) consider thermodynamics to be fundamental because one can be extremely sure that it holds. In contrast, in the quote from \citet{TadmorME2012} (which comes from an engineering textbook), \ZT{fundamental theories} presumably means something much more modest, along the lines of \ZT{the most important theories for science and engineering} or \ZT{the theory that most of science and engineering are based on}. A reasonable debate about whether thermodynamics is fundamental calls for some account of what we take \ZT{fundamental} to mean.
	
	There are several notions of fundamentality in the philosophical literature. Foundationalist conceptions hold that there is a fundamental level of reality, usually conceived as inhabited by a realm of entities (\textit{pluralism}) rather than a single one, and/or characterised by a range of facts not depending on anything else at the base of a hierarchy, upper levels of which depend on lower. But foundationalism may also take the form of monism, according to which there is a single fundamental thing, the universe as a whole, whose parts depend on the whole and which could generate a hierarchy of dependent parts \citep{Morganti2020,Morganti2020b}. Although such hierarchies express an ordering of what is more fundamental than what, distinguishing more fundamental entities or facts from others need not involve commitment to either form of fundamentalism---whether there are ultimate atoms or an all-encompassing whole. Often, the more fundamental relation is understood based on grounding \citep{Tahko2018}. Grounding, however, is an ontological relation that holds between things in the world (objects or facts), but not between descriptions of the world such as scientific theories.
	
	The more fundamental relation might, in a scientific context, be construed in terms of reduction, what is less fundamental being reducible to what is more fundamental, and this in turn might be construed as being derivable from, in some appropriate sense of \ZT{derivable}. Derivability need not necessarily be understood in the now unfashionable Nagelian sense \citep[Ch. 11]{Nagel1961} of being logically deducible from, but more loosely, allowing for approximations and for conclusions falling within specified margins of error and under certain restrictions. Although derivations may take very different forms in particular cases, they generally fall back on a well-developed understanding of the appropriate requirements, lack of which is the main criticism of the notion of grounding \citep{BlissT2021}. In the wake of criticisms of the inapplicability of the Nagelian requirement of strict logical deduction, doubts have even arisen about the feasibility of general models of reduction which can serve as criteria by which particular cases of reduction might be reasonably assessed. \citet{Crowther2020} seems to take the relation of being more fundamental between theories, or parts thereof, as given and understands reduction in terms of being broadly derivable (within appropriate limits of precision) from a more fundamental theory.
	
	Questions of fundamentality are concerned with objects or what is said of objects with the aid of predicates expressing properties and relations used to formulate facts with sentences conjoined in the form of theories. Here our main concern is with theories and the properties and relations they govern, without commitment to any form of foundationalism or hierarchy of facts or objects. From this point of view, an appropriate notion of fundamentality is that a fact or theory is fundamental if it is part of or belongs to a complete minimal basis (CMB) that determines everything else or to which everything else is reduced\footnote{We here apply the idea of a CMB to scientific theories. Note that other authors, such as \citet{Bennett2017}, have used it for understanding the fundamentality of entities instead.}. “Complete” here indicates that everything else is reduced to the CMB, and “minimal” means that no proper part (subset) of this basis is complete \citep{Tahko2018}. Ideally, a consistency condition should be satisfied to preclude the trivial case of a contradiction entailing anything. Perhaps this is entailed by the condition that no proper part is complete. But as acknowledged in \cref{general}, in the present state of physics there are plausible candidates for membership of a CMB that are mutually inconsistent. Since we are not dealing here with some unspecified ideal future state of science, a provisional condition would be to consider a fact to be accommodated by the CMB only if it is accommodated by a consistent part of it, and the minimality condition is that the system would not be complete were a proper part removed from any consistent part. This seems close to the way in which physics currently looks for a fundamental theory, namely by looking for a small set of principles that everything else can be derived from. 
	
	A particular axiomatization of a theory would take certain predicates as primitive and define others in terms of them. Certain extensive properties might be taken as primitive in a particular formulation of thermodynamics, for example, and intensive properties defined in terms of them. But another formulation might adopt some of the defined terms of the first formulation as primitive and introduce definitions of what were primitives in the first formulation. Clearly, being a primitive predicate in a particular formulation of a theory does not make it more fundamental than the predicates defined in terms of it. On the CMB conception of fundamentality, all predicates of a theory in the CMB would have the same status. Predicates featuring in non-fundamental theories reducible to the CMB would be non-fundamental only if eliminable by reduction to predicates featuring in fundamental theories in the CMB.
	Here we take it that thermodynamics is fundamental, i.e. part of some CMB to which everything is reducible. This article will be concerned for the most part with objections to the reducibility of thermodynamics based on the idea of multiple realizability. A discussion of other objections against the fundamentality of thermodynamics is taken up later, in \cref{other}.
	
	The reductionist view on science \citep{OppenheimP1958}, and with it the idea that the fundamental theories are those that everything else is reduced to, has fallen out of fashion in recent years. Neither has physics been able to (or can be expected to be able to in the next few years) find a \ZT{theory of everything}, nor has the reduction of most (if any) special sciences to physics been achieved in practice. For this reason, \citet{Morganti2020b} gives, in his review of fundamentality in philosophy of physics, this idea only a rather brief mention before moving to more sophisticated concepts. For example, one may assume that there is something like a \ZT{layered reality} and then take the fundamental theory to be the description of the deepest layer, or one may (as suggested by \citet{Ney2019}) define fundamentality in terms of \textit{explanatory maximality} rather than explanatory completeness, i.e., the fundamental theory is the one that has the largest scope and accuracy.
	
	However, as long as one does not reject the idea of reduction altogether (and given that reduction is still widely discussed in philosophy of science, this does not appear to be a majority view), it is still reasonable to argue that, if a theory $A$ can be reduced to another theory $B$, theory $B$ is more fundamental than theory $A$, and that a fundamental theory should not be reducible to another one. Hence, rather than giving up the idea of a CMB in general, one can adapt it to anti-reductionist challenges by refining the notion of \ZT{everything} appearing in its definition. If, for example, psychology or the social sciences cannot, in principle, be reduced to physics, then the completeness criterion might be satisfied already by a set of axioms that everything \textit{in physics} can be reduced to. The notion of fundamentality would thereby become a more context-dependent one, but this is quite in line with how the understanding of fundamentality has recently developed \citep{Morganti2020b}.
	
	\section{\label{ramr}Reduction and multiple realizability}
	We now turn to a particularly interesting issue which may pose a problem for reduction, namely \textit{multiple realizability}. The idea of multiple realizability as a problem for reduction first appeared as an argument in the philosophy of mind in opposition to the view (defended, e.g., by \citet{Place1956} and \citet{Smart1959}) that every mental kind is identical with a neural kind, so that pain is identical with a certain brain state. Against this, \citet{Putnam1967} has argued that many different animals---humans, reptiles, birds, ...---all seem to experience pain, but have quite different neuroanatomies, which makes it unlikely that they all exhibit the same neural kind. Moreover, according to the identity thesis, the identity between pain and neural state would be a natural law, so that anything feeling pain would exemplify this neural kind. But while it seems plausible that Martians or sentient robots could feel pain, their neuroanatomy would be very different from that of humans, which makes it implausible that they would exhibit the same neural kinds. There is a distinction here between multiple \textit{realizability} (there \textit{could} be different realizations) and multiple \textit{realization} (there \textit{are} different realizations). Pain in humans and reptiles is an example of multiple realization, whereas pain in humans and Martians is an example of multiple realizability.
	
	\citet{Fodor1974} takes reductionism to imply that every natural kind in a special science corresponds to a natural kind in physics (assuming both physics and the special science to be ideally completed). Reductionism would then entail that there is a natural kind in physics that corresponds to pain. But due to multiple realizability, what corresponds to the mental kind \ZT{pain} in physics would be a disjunction of all possible physical realizations of pain. This disjunction will (due to the diversity of the possible realizations) not be a natural kind figuring in any physical theory. Consequently, there is no natural kind in physics that corresponds to the mental kind \ZT{pain}, and thus reductionism fails. (This paragraph and the previous paragraph follow \citet{Bickle2020}.)
	
	The problem of multiple realizability also arises in the context of thermodynamics. Although the statement \ZT{temperature is mean molecular kinetic energy} has been taken as a paradigmatic example of reduction, this is unfortunate since the claim is, for most systems, wrong \citep{MenonC2013}. Electromagnetic radiation or a system of spins can also have a temperature (playing the same physical role as the temperature of gases), yet this temperature cannot be interpreted as a mean kinetic energy \citep{Needham2010}. A more extensive discussion of multiple realizability in the context of thermodynamics/statistical mechanics was provided by \citet{Batterman2000} using the example of renormalization group theory. 
	
	Note that by \ZT{temperature} we always mean the thermodynamic temperature here. One can, of course, also \textit{define} temperature as mean molecular kinetic energy; this is referred to as \ZT{kinetic temperature} \cite{MandalLL2019}. If we speak about temperature in thermodynamics, we do not usually mean the kinetic temperature since the thermodynamic temperature of a system is not usually given by the kinetic temperature (except, e.g., for a gas in thermal equilibrium). Similar ambiguities can arise also for other quantities, in particular for active matter systems (which consist of self-propelled particles and are far from thermodynamic equilibrium). Here, it has been found that, e.g., the mechanical definitions of pressure \citep{WittkowskiTSAMC2014} or flow velocity \citep{teVrugtFHHTW2022} do not coincide with the thermodynamic definitions of the same quantities. This, of course, makes it even clearer that, e.g., the thermodynamic pressure is not simply identical with some mechanical quantity, otherwise it would not be possible for them to disagree. 
	
	There is no universal agreement that multiple realizability really constitutes a problem for reduction. \citet{Clapp2001} has argued that it is legitimate to identify higher-level kinds with disjunctive kinds, while \citet{Kim1992} argues that while kinds might have multiple realizations, they have only one in each specific context. For example, human-pain or reptile-pain correspond to only one brain state, i.e., one neural kind. We thus have a series of \ZT{local reductions} \citep{Kim1989}, and this is all that we can reasonably ask for. Temperature, according to this view, cannot be reduced in general, but can be reduced for each system it plays a role in. Regarding such properties, \citet[pp. 17-18]{Kim1999} argues:\footnote{In this quote, $E$ is a multiple realizable property.}
	\begin{quotation}
		It follows then that multiply realizable properties are ipso facto causally and nomologically heterogeneous. This is especially obvious when one reflects on the causal inheritance principle. All this points to the inescapable conclusion that $E$, because of its causal/nomic heterogeneity, is unfit to figure in laws, and is thereby disqualified as a useful scientific property
	\end{quotation}
	However, at least for the case of temperature this response does not do justice to the corresponding science (here thermodynamics) \citep{Needham2009}. A box of gas (where temperature is realized by mean kinetic energy) can be in thermal equilibrium with a radiation heat bath (where temperature is realized by the temperature of the electromagnetic radiation). If we distinguish between kinetic temperature and radiation temperature and just reduce each of them, we are overlooking the fact that the reason that box and radiation are in equilibrium is precisely the fact that they exhibit one common property, namely that of having the same temperature (regardless of which microscopic origin this temperature has). An account of reduction that makes temperature \ZT{unfit to figure in laws} is, in our view, unfit to provide a reasonable account of physics. 
	
	The fact that thermodynamics has figured as a paradigm example in the discussion has led to it playing a central role in the debate concerning psychophysical reduction. \citet{Bickle1998} has used it in his defense of the so-called \ZT{new wave} reduction to block what he views as the strongest type of multiple realizability argument, namely one based on multiple realizability within the same individual:
	
	\begin{quotation}
		Consider temperature (as a thermodynamic kind). For any aggregate of molecules, there are an \textit{indefinite} number of realizations of the kinetic-theory analog, \textit{mean} molecular kinetic energy, in terms of the \textit{microcanonical ensemble}, the finest specification of a gas's microphysical state in which the location and the momentum (and thus the kinetic energy) of each constituent molecule are individually fixed. Obviously, indefinitely many distinct microcanonical ensembles of a fixed volume of gas can produce the same \textit{mean} molecular kinetic energy. Thus at the lowest level of microphysical description, temperature is vastly multiply realizable in the same aggregate of molecules at different times. Yet this reduction is a textbook example of intertheoretic reduction, reconstructable within the new-wave account. \citep[p. 125]{Bickle1998}
	\end{quotation}
	
	This discussion confuses \ZT{microcanonical ensemble} and \ZT{point in phase space}\footnote{A point in phase space is determined by positions and momenta of all particles and fully specifies the microstate of a classical many-body system. The microcanonical ensemble is a probability distribution over phase space that assigns the same probability to all states with a certain total energy $E$ and zero to all others. For an ideal gas with $N$ particles (where $E$ is just the total molecular kinetic energy), there is thus one microcanonical ensemble with a certain mean molecular kinetic energy $E_M$---namely that with total energy $NE_M$---but this ensemble allows for infinitely many microstates with the same mean molecular kinetic energy.}, but the idea is clear enough. Essentially, Bickle argues that multiple realizability cannot be a problem for psychophysical reductionism because it can also be found in thermodynamics, which is reducible. We appear to have an argument of the following form
	\begin{itemize}
		\item Premise 1: Thermodynamics is reducible to statistical mechanics.
		\item Premise 2: If reduction does not allow for multiple realizability, then thermodynamics is not reducible to statistical mechanics.
		\item Conclusion: Reduction allows for multiple realizability.
	\end{itemize}
	As an argument against the multiple realizability objection against reduction, this of course begs the question---perhaps thermodynamics is not reducible (because of multiple realizability) despite the widespread belief that it is. \citeauthor{Bickle1998} \citeyearpar[pp. 33-40]{Bickle1998} provides an explicit discussion of how thermodynamics is supposedly reduced to statistical mechanics, and thus takes it as an established fact that premise 1 is true. However, his analysis of the reduction focuses exclusively on gases and thus does not solve the problem in general. A more generous interpretation of Bickle’s argument would be that the reducibility of thermodynamics is used as a premise in an argument countering the multiple realizability objection against psychophysical reduction. But then, this premise calls for further investigation. Due to the central role thermodynamics plays in the reduction debate, the problem of the multiple realizability of temperature is thus also of wider interest in philosophy of science.
	
	A similar point of view has also been expressed by \citet[p. 200]{MenonC2013}:
	\begin{quotation}
		One does not need to look at phase transitions to notice that any claim about the reduction of thermodynamics to statistical mechanics must be based on a conception of reduction that is compatible with multiple realizability. Temperature, that most basic of thermodynamic properties, is not (the claims of numerous philosophers notwithstanding) simply \ZT{mean molecular kinetic energy}. It is a multiply realizable functional kind. If our notion of reduction precludes the existence of such properties, then the project of reducing thermodynamics cannot even get off the ground.
	\end{quotation}
	Of course, we fully agree with the observation that temperature exhibits multiple realizability. However, the wording in the quote from Menon and Callender (and also the discussion following it) strongly suggests that they see \ZT{the project of reduction not getting off the ground} as something problematic that should be avoided. In our opinion, however, the analysis should go exactly the other way round---multiple realizability poses a problem for reduction, and if this means that a specific project of reduction cannot get off the ground, then we have to live with it.
	
	At this point, we should discuss the following response to our arguments against Bickle, Menon, and Callender: The adequacy of a definition of \ZT{reduction} should take into consideration whether it classifies things as \ZT{reduction} that we think should clearly be classified as such. After all, we would not have an idea of what reduction probably is without such examples. (Similarly, a definition of \ZT{science} which excludes physics as a science or counts homeopathy as a science would be reasonably considered a failure because we are sure that whatever \ZT{science} might precisely be, physics surely is one and homeopathy surely is not.) We might now proceed to argue that the relation of thermodynamics and statistical mechanics is such a clear case, and that we should therefore adapt our analysis of reduction in such a way that it classifies the relation between thermodynamics and statistical mechanics, whatever else it might be, as reduction.
	
	Two arguments can be raised against this objection. First, it is not consistent with the way that the problem \ZT{Can thermodynamics be reduced to statistical mechanics?} is generally discussed, given that a variety of authors have expressed considerable scepticism regarding the reducibility of thermodynamics. Thus (unlike in the cases of physics and homeopathy) we do not appear to have a \ZT{clear case} here. Second, it should be noted that, even though we spend most of this article talking about reduction, we are ultimately interested in fundamentality. In our view, the derivability of thermodynamics from statistical mechanics faces so many problems that, if one adapts the definition of \ZT{reduction} in such a way that the relation between thermodynamics and statistical mechanics still counts as a reduction, then this notion of \ZT{being reducible to} is not an appropriate basis for a definition of \ZT{being more fundamental than}.

	\section{\label{entropicgravity}Entropic gravity: a case study for multiple realizablity}
	So far, we have addressed the fundamentality and reducibility of thermodynamics in a rather general way. Now we present two arguments against the reducibility of thermodynamics to statistical mechanics that are both based on the multiple realizability problem. The first concerns entropic gravity, the second classical black hole thermodynamics.
	
	Let us start with a brief recap the essential ideas behind the derivation of the Newton's law of gravity in the framework of entropic gravity by \citet{Verlinde2011}. The idea of entropic gravity is to show that (three-dimensional) space and gravity are emergent. It is based on the holographic principle, a standard result from quantum gravity according to which all three-dimensional systems can be described simply via information encoded on their two-dimensional surface (by analogy with a hologram, where a three-dimensional picture is saved on a two-dimensional surface). Depending on one's physical and philosophical commitments, the surface description can be considered more fundamental than the bulk description, and thus the holographic principle can be taken to show that the world, on a fundamental level, is two- and not three-dimensional \citep[pp. 204-205]{DieksvDdH2015}. 
	
	What is conceptually interesting here is that gravity is taken to be an \ZT{entropic} force. Such forces arise not from fundamental interactions, but for statistical reasons. A well-known example (also used by \citet{Verlinde2011} as an illustration) is a polymer. If we stretch it by an amount $\Delta x$, the number of possible configurations (and therefore the entropy) is reduced. The polymer twists in order to maximize its entropy---an effect that you will be familiar with, e.g., from cables on your headphones or computer. Consequently, by its tendency to approach equilibrium (state with maximal entropy), the polymer will exhibit an \ZT{entropic} force counteracting the force that is pulling it. This force is governed by expressions of the form \eqref{eg6} (see below). The basic idea behind entropic gravity is that gravity is nothing more than this---the particle's motion in a \ZT{gravitational field} corresponds to the tendency of the bits of information on a surface to maximize their entropy.
	
	We follow \citet{Verlinde2011} and \citet{DieksvDdH2015}. Let us consider a spherical surface on which some unspecified microscopic dynamics takes place. This dynamics can be described in an information-theoretical way, according to which little pieces of this surface encode a bit of information. These bits tend towards a thermodynamic equilibrium distribution. The number of bits $N$ on this surface is
	\begin{equation}
	N=\frac{c^3}{G\hbar}A.
	\label{eg1}
	\end{equation}
	
	Here, the reasonable assumption has been made that $N$ is proportional to the area $A$, and the proportionality constant has been written as $c^3/(G\hbar)$. This serves as a definition of $G$, while $c$ is the speed of light and $\hbar$ the reduced Planck constant. We then define a radius $R$ by
	\begin{equation}
	A=4\pi R^2,
	\label{eg2}
	\end{equation}
	and introduce a temperature $T$ via the equipartition rule
	\begin{equation}
	E=\frac{1}{2}Nk_B T,
	\label{eg3}
	\end{equation}
	where $E$ is the total energy, assumed to be distributed evenly over all degrees of freedom, and $k_B$ is the Boltzmann constant. The energy can be related to a mass $M$ by
	\begin{equation}
	E=M c^2.
	\label{eg4}
	\end{equation}
	
	Via the holographic principle, the dynamics of the bits on the surface can be considered to describe the dynamics of things in what might prove to be the \ZT{inside} of the surface. Now assume that, on this inside, a particle with mass $m$ approaches the screen by a distance $\Delta x$. Motivated by an argument from \citet{Bekenstein1973}, Verlinde takes this to correspond to an entropy change
	\begin{equation}
	\Delta S = \frac{2\pi k_B m c}{\hbar}\Delta x.
	\label{eg5}
	\end{equation}
	We finally define an \ZT{entropic force} by
	\begin{equation}
	F\Delta x = T\Delta S.
	\label{eg6}
	\end{equation}
	Equations \eqref{eg1} -- \eqref{eg6} can be combined to give
	\begin{equation}
	F = \frac{G M m}{R^2}.
	\end{equation}
	
	In summary, microscopic dynamics describing the information on the holographic screen gives rise to thermodynamic behaviour on a large-scale level, which leads to the emergence of gravity. If this theory is correct (which, of course, is a very big if), then we can construct a rather straightforward argument for why thermodynamics is more fundamental than gravity: Entropic (i.e., thermodynamic) effects are the reason for the existence of gravity, and thus general relativity can be reduced to thermodynamics. 
	
	However, entropic gravity does not, prima facie, appear to provide an argument for the fundamentality of thermodynamics. Recall that Verlinde's argumentation assumes gravity to be related to coarse-graining, and thus to emerge from some microscopic dynamics. In this sense, it is this microscopic dynamics that is most fundamental. However, Verlinde's article also makes little to no explicit assumptions as to what this microscopic dynamics looks like. It is simply assumed that whatever it is, it will give rise to the sort of relations between (for example) entropy and temperature that we are familiar with from more \ZT{standard} applications of thermodynamics (such as in polymer physics). Let us call the actual microdynamics M$_1$, and let us assume that M$_2$,...,M$_n$ are some other microdynamics that also give rise to thermodynamics on a higher level. Gravity would then emerge also from, say, M$_2$. Thus, gravity is not explained by the fact that the microdynamics is M$_1$ rather than M$_2$. A CMB in the sense of a \ZT{theory of everything} to which the laws governing all physical interactions \textit{including gravity} can be reduced therefore needs to include thermodynamics.  
	
	By the general multiple realizability argument, this would then imply that, while we perhaps can reduce gravity to thermodynamics, we \textit{cannot} reduce it to $M_1$. Gravity would here be analogous to pain: Gravity exists in a universe with microdynamics $M_1$ and exists in a universe with microdynamics $M_2$, just like pain exists in humans, reptiles, and Martians with different physiological realizations. If pain is, due to this multiple realizability, not reducible to the microscopic physiology of humans, reptiles, and Martians, then gravity is not reducible to $M_1$ (or $M_2$), although it is (or at least might be) reducible to thermodynamics.
	
	An objection that comes to mind is that we are dealing here with \textit{hypothetical} different microdynamics. Pain in humans and pain in reptiles both exist, whereas in the gravity case only $M_1$ is the actual microdynamics. We are, to use the terminology introduced in \cref{ramr}, concerned with multiple realizability rather than multiple realization. However, it is not clear why this should be detrimental to the force of the anti-reductionist argument. As far as we currently know, Martians do not exist (let alone their pain), but nevertheless the pain of Martians plays a considerable role in the debate about psychophysical reductionism. Second, imagine a situation in which suddenly all animals except for the humans are killed (for example because the earth is destroyed and only the humans escape via a spaceship). If pain was not reducible prior to this catastrophe because it existed (with a different physiological realization) in reptiles, then it seems unlikely that it suddenly becomes reducible after the last crocodile has stopped breathing. Thus, reducibility does not appear to depend on whether possible alternative microrealizations actually do exist at the time in question.
	
	Finally, one might wonder why we have posed the argument in this section using the rather speculative example of entropic gravity rather than based on some other case where it is established (rather than just assumed) that thermodynamics can emerge from different microdynamics. There are two reasons. First, general relativity is usually taken to be a fundamental theory, and thus a relation of thermodynamics to general relativity is of particular interest for the problem of fundamentality. Second, by analyzing the case of entropic gravity, we have not only learned something about the reduction of thermodynamics; we have also learned something about the (possible) reduction of general relativity.
	
	\paragraph{}
	\section{\label{cbht}Classical black hole thermodynamics: a second case study in multiple realizability} 
	A second argument for the multiple realizability thesis in the context of gravity and thermodynamics can be obtained from \textit{black hole thermodynamics} \citep{Bekenstein1972}. This field is interesting since---as will be explained below---it provides (in its classical form) a possible case in which entropy and temperature are not related to any statistical theory.
	
	Black hole thermodynamics emerged from the observations that the laws  governing the mechanics of black holes \citep{BardeenCH1973} can be stated in a way that has a striking similarity to the laws of thermodynamics:
	\begin{itemize}
		\item Zeroth law: Throughout a black hole's event horizon, the surface gravity $\varkappa$ is constant. (Analogous to the zeroth law of thermodynamics requiring constancy of $T$ throughout a system in thermal equilibrium.)
		\item First law: The total mass $M$, its area $A$, rotational velocity $\Omega$, and angular momentum $J$ are related by 
		\begin{equation}
		\dif M = \frac{\varkappa}{8\pi}\dif A + \Omega \dif J.    
		\end{equation}
		(Analogous to the first law of thermodynamics requiring $\dif E = T\dif S - p\dif V + \Omega \dif J$).
		\item Second law: For any process, one has $\dif A \geq 0$. (Analogous to second law of thermodynamics, requiring $\dif S \geq 0$).
		\item Third law: $\kappa = 0$ cannot be reached. (Analogous to the third law of thermodynamics, requiring that $T=0$ cannot be reached).
	\end{itemize}
	(List adapted from \citet{Curiel2014}.) Note that the laws of black hole thermodynamics are purely geometrical results (more specifically, theorems of differential geometry \citep{DoughertyC2016}). As can be seen, there is a quite remarkable analogy between the laws governing black holes and the laws of ordinary thermodynamics. In these laws, the surface gravity $\varkappa$ plays the role of the temperature and the area $A$ plays the role of an entropy.
	
	There has been a considerable debate in philosophy of physics about whether this is more than an analogy. \citet{DoughertyC2016}, for example, have argued that this is in fact nothing more than a formal analogy, whereas \citet{Wallace2018} has argued that black holes are genuine thermodynamic objects. A widely held view is something like the following \citep{Curiel2014}: For classical black holes, the surface gravity is not a temperature (the analogy between classical thermodynamics and the laws for black holes is merely formal) since they cannot come into thermal equilibrium in the way an ordinary thermodynamic object would---the latter would, if placed in a heat bath of thermal radiation with temperature $T$, itself (in equilibrium) emit radiation with such a temperature and therefore \textit{have} this temperature. A classical black hole, by contrast, would simply absorb all radiation and therefore have a temperature of zero. Therefore, its surface gravity has nothing to do with a temperature. A \textit{quantum} black hole, on the other hand, would emit Hawking radiation with temperature $\hbar\varkappa/(2\pi)$, such that the surface gravity is (up to a prefactor) just the temperature defined in the usual way. For quantum black holes, the analogy is therefore more than formal.
	
	\citet{Curiel2014} has argued against this view by pointing out that defining heat exchange via an equilibrium in terms of electromagnetic radiation might not be appropriate for classical black holes, which will emit energy via \textit{gravitational} radiation (a phenomenon that has recently been confirmed experimentally \citepalias{Abbott2016}). Just as electromagnetic radiation was found to be a mode of heat exchange, so might gravitational radiation. Curiel then proceeds by constructing a Carnot cycle in which a classical black hole couples to an ordinary thermodynamic system. In this cycle, the black hole's surface gravity plays the role of temperature and the area plays the role of entropy.
	
	Suppose Curiel's analysis is right\footnote{There is no room here for an extended discussion of whether it actually is, so everything we write in this section should be understood as an analysis of what Curiel's analysis would imply for our question if it were correct.}. In this case, we would have another argument for the multiple realizability thesis. The reason is that the statistical mechanics of black holes is based on quantum gravity \citep{Wallace2019}, whereas classical black holes are extremely simple objects that can be fully described by a very small number of parameters. Consequently, the entropy of a \textit{classical} black hole does not---unlike the entropy of a classical many-particle system---arise from a large-scale statistical description, it is simply a basic property of the black hole, which is a single object. 
	
	It seems that at this point, our position could be subject to a similar criticism to that aimed at \citeauthor{Lewis1986}' \citeyearpar{Lewis1986} discussion of relations, which involved using the example of a classical hydrogen atom. Against this, \citet[p. 20]{LadymanR2007} argue that \ZT{[t]here are not, nor were there ever, any ‘classical hydrogen atoms’}. Their broader message is that philosophical inferences often start from outdated scientific views, and that this often leads to unreasonable metaphysical positions. If we now develop a philosophical argument based on a thermodynamic perspective on classical general relativity, aren't we making the same mistake given that the world is fundamentally quantum?
	
	First, it should be noted that, while there is a very well established quantum theory of the hydrogen atom, there is no established quantum theory of gravity and one could also construct such an argument in precisely the opposite way (classical general relativity is the best empirically confirmed theory we have in this field and should thus be our starting point). Second, even if we assume that black holes have a quantum description like the one used in (for example) string theory, this does not change the fact that even if they did not have such a quantum description (i.e., if they were completely classical objects), they would still be subject to thermodynamics. Consequently, the fact that they can be described by thermodynamics does not hold \textit{in virtue of} them having a statistical-mechanical description, and is thus not less fundamental. On the \ZT{quantum viewpoint}, we are, to use the terminology from \cref{ramr}, concerned with multiple realizability and not with multiple realization (temperature can be realized both by mean molecular kinetic energy and by the surface gravity of classical black holes).
	
	Classical black holes thus provide a particularly interesting example of multiple realizability. If the surface gravity of a classical black hole is a physical temperature, then it would provide a particularly striking counterexample for the typical claim that \ZT{temperature is mean molecular kinetic energy}. The black hole has no constituents that could have a kinetic energy (in fact, it has no constituents at all since it is just a region of spacetime), yet it still has a temperature. We have thus found a realization of \ZT{temperature} that is very different from \ZT{mean kinetic energy}, further supporting the multiple realizability thesis regarding thermodynamics. While the fact that there are different ways to realize \ZT{temperature} microscopically in different systems (such as ideal gases, solids, radiation baths, ...) has been acknowledged by a variety of authors \citep{Needham2010,Bickle1998,MenonC2013}, it is quite remarkable to find a realization that does not have anything to do with statistical mechanics, i.e., that is not \ZT{microscopic} at all.
	
	One could now, following \citet{Kim1999}, infer from the multiple realizability of temperature that we should not think about \ZT{temperature} as a property that can play a role in scientific explanations (see \cref{ramr}). However, if we wish to explain why a black hole and a block of iron in thermal contact exchange no heat, then the explanation is the fact that they have the same temperature. For the role an object plays in a thermodynamic cycle, it does (at least on an abstract level) not matter whether it has a certain mean molecular kinetic energy or a certain surface gravity as long as these lead to the same temperature. See \citet{Needham2009} for a similar response to Kim in the context of thermodynamics and \citet{Marras2002} for such a response in the context of psychology.
	
	Another approach that is worth mentioning here is the derivation of equations for gravity using the phase-field formalism by \citet{Schmitz2017}. (Phase fields are discussed in more detail in \cref{pt}.) This allows, in a purely thermodynamic way based on a combined phase-field/entropic approach, the derivation of equations with the same structure as those for gravity. The phase-field here can be considered as \ZT{spatially resolved thermodynamics}. This formalism leads to the Poisson equation of gravitation and also includes terms which are related to the curvature of space. It further generates terms possibly explaining nonlinear extensions known from  modified Newtonian dynamics (MOND) \citep{Milgrom1983} approaches. As an example, consider the dimensionless entropy of a black hole given by
	\begin{equation}
	S=\frac {A}{4l_p^2}  
	\label{bhentropy}
	\end{equation}
	with the Planck length $l_p$. Equation \eqref{bhentropy} can be derived within the phase-field approach when using it to assign an entropy to a geometric object (in this case a sphere) \citep{Schmitz2018}. See \citet{SchmitztVHWENW2022} for a further development of this idea. This is another example where thermodynamic considerations are used even though there is no microscopic physics that one could do statistical mechanics for in order to derive it (a geometric sphere does not consist of particles).
	
	In summary, we have, using the examples of entropic gravity and classical black hole thermodynamics, shown that thermodynamic aspects of gravity provide an interesting argument in favour of the multiple realizability thesis regarding thermodynamics, and thus against the reducibility of thermodynamics to statistical mechanics. 
	
	\section{\label{pt}Phase transitions and the phase-field model}
	A widely discussed problem in the debate about the alleged reduction of thermodynamics is the definition of phase transitions. Thermodynamics describes phase transitions using singularities of a thermodynamic potential such as the free energy.\footnote{The order of the discontinuous derivative is a basis for classifying phase transitions in the traditional Ehrenfest scheme, but this scheme has been replaced in modern thermodynamics and is not appropriate for the types of phase transitions discussed later in this section. See \citet{Jaeger1998} for a historical discussion.} Using the microscopic definition of the free energy based on the partition function $Z$ from statistical mechanics, however, one can show that the free energy cannot have a singularity except in the thermodynamic limit where the particle number $N$ goes to infinity \citep{MenonC2013}.  \citet[p. 549]{Callender2001} summarizes this problem in the following propositions
	\begin{quotation}
		\begin{enumerate}
			\item real systems have ﬁnite $N$;
			\item real systems display phase transitions;
			\item phase transitions occur when the partition function has a singularity;
			\item phase transitions are governed/described by classical or quantum statistical mechanics (through $Z$)
		\end{enumerate}
	\end{quotation}
	These four propositions are inconsistent because in statistical mechanics, $Z$ can only have a singularity for infinite $N$. Various authors, such as \citet{PrigogineS1997} and \citet{Liu1999}, have rejected proposition 4 and inferred that phase transitions are an emergent, non-reducible phenomenon. If proposition 4 is incorrect, then statistical mechanics cannot describe phase transitions (a central phenomenon in thermodynamics), and thus thermodynamics cannot be reduced to statistical mechanics. As emphasized by \citet{MenonC2013}, this problem is closely related to the issue of multiple realizability. From the perspective of finite-$N$ statistical mechanics, there is only a quantitative difference between the transition from water to ice and the transition from cold to lukewarm water. Consequently, statistical mechanics is unable to identify the various phenomena which we refer to as \ZT{phase transitions} as a single natural kind.
	
	\citet[p. 550]{Callender2001}, however, argues
	\begin{quotation}
		Clearly the weakest link in the chain is 3; consequently, we ought to affirm 1, 2, 4 and conclude the denial of 3. That is, we should say that real ﬁnite systems give rise to the sort of behaviour associated with phase transitions in thermodynamics even when the partition function is not singular.
	\end{quotation}
	What Callender wishes to show here is that the conventional argument for the non-reducibility of thermodynamics is based on taking a particular aspect of thermodynamics (the fact that it describes phase transitions by singularities) too seriously. Denying proposition 3 is an alternative way of avoiding the inconsistency illustrated above, and one that is supposed to leave the reduction of thermodynamics to statistical mechanics unharmed. 
	
	However, denying proposition 3 leaves the problem of identifying a natural kind \ZT{phase transitions} unsolved, since abandoning  3 means we now no longer have a definition of a phase transition. Several solutions to this problem have been proposed \citep{MenonC2013}. First, one could argue that we were wrong to think about phase transitions as a natural kind---perhaps phase transitions are, as \citet[p. 778]{Kadanoff2009} argued, just \ZT{products of the human imagination}. Second, one could try to re-define phase transitions in such a way that they can arise already in finite $N$ systems. Various ways to do this have been proposed (see \citet{MenonC2013}). In both cases, we would have a correction of thermodynamics by statistical mechanics, which can be advanced as an argument against seeing thermodynamics as fundamental. And in both cases, this is based on the assumption that thermodynamics is not capable of modelling finite-size effects. In our view, this is a consequence of the fact that philosophers \textit{don't take thermodynamics seriously enough}. Modern approaches of thermodynamics are considerably more powerful in this regard than philosophers usually believe. To see why, it is helpful to take a brief look at some limitations of the classical approach and treatments of phase transitions in modern materials science.
	
	While classical thermodynamics allows for the description of phase-equilibria between bulk phases, it does not describe the process of a phase transition itself. The reason is that this would also require the incorporation of the existence of boundaries between different phases which appear and disappear during the phase transition. This is beyond the scope of classical thermodynamics, which does not have a specific length scale and so does not provide information about structures that do have length scales. Describing the evolution of structures requires the description of interfaces, and the finite thickness of interfaces introduces a characteristic length scale \citep{Schmitz2003}. The basic premise of \textit{phase-field models} is the existence of a diffusive interface between two phases, which is described by a phase field $\phi$ \citep{Steinbach2009}. Such an interface can be introduced simply for numerical convenience, but also because its existence is implied by basic principles of irreversible thermodynamics pioneered by \citet{Onsager1931} \citep{Emmerich2008}. Early models of this type are the \textit{Cahn-Hilliard equation} \citep{Cahn1965,CahnH1958} and the \textit{Allen-Cahn equation} \citep{AllenC1976}, which describe a conserved and a non-conserved order parameter $\phi$, respectively.
	
	Phase field models, which are nowadays regularly applied to describe phase transitions in technical processes like solidification of complex alloys, thus constitute an extension of thermodynamics towards \ZT{spatially resolved thermodynamics} or \ZT{time-dependent theories of phase transitions}. In the materials science community, phase-field models, in particular their numerical implementation in the form of software packages such as \citet{Micress} have been widely adopted because of their ability to describe phase transitions in complex alloy systems and the resulting evolution of highly complex structures relevant for technical applications. Moreover, the phase-field concept itself can be related to fundamental concepts of analytical philosophy like mereology \citep{Schmitz2020} and mereotopology \citep{Schmitz2022}. Reviews of the phase-field approach can be found in \citet{Emmerich2008}, \citet{Steinbach2009}, and \citet{ProvatasE2011}.

	The thermodynamic limit has been studied in recent work by \citet{ThieleFHEKA2019} using the example of liquid-liquid phase separation, which can be described using the Cahn-Hilliard equation. This equation is given by
	\begin{equation}
	\partial_t\phi = \Nabla\cdot\bigg(Q(\phi)\Nabla\Fdif{F}{\phi}\bigg)
	\label{cahnh}
	\end{equation}
	with a phase-field $\phi$, a positive mobility $Q$, and a free energy
	\begin{equation}
	F=\INT{}{}{}\frac{\kappa}{2}(\Nabla\phi)^2+\frac{\tilde{a}}{2}\phi^2+\frac{b}{4}\phi^4,   
	\label{freeenergy}
	\end{equation}
	where $\kappa$, $\tilde{a}$, and $b$ are positive constants. In particular, we have $\tilde{a}=a(T-T_c)$ with the temperature $T$, the critical temperature $T_c$, and $a>0$. The Cahn-Hilliard theory allows us to find the equilibrium state of the system as the minimum of a free energy that is a very simple function of a local concentration field $\phi$. In the thermodynamic limit of an infinite system size (and only there), the critical temperature at which a phase transition occurs can be found using a simple standard procedure (known as the Maxwell construction). Here, the order parameter changes discontinuously. To see this, note that the gradient contribution in \cref{freeenergy} (the term proportional to $\kappa$) becomes irrelevant for an infinite system, so that the condition for a minimum of $F$ (equilibrium) becomes
	\begin{equation}
	\mu = a(T-T_c)\phi + b\phi^3    
	\end{equation}
	with the constant chemical potential $\mu$. For $T>T_c$, there is only one solution (the homogeneous density $\phi_0$), whereas for $T<T_c$, there can (depending on $T$ and $\phi_0$) be demixing in regions with densities $\phi_\pm = \pm\sqrt{a(T_c-T)/b}$ (binodals). The homogeneous state $\phi_0$ remains metastable when crossing the binodal and becomes unstable after crossing the spinodal (where it changes from a local minimum to a local maximum of the free energy). The spinodals are given by $\phi_s = \pm\sqrt{a(T_c-T)/(3b)}$. Since phase coexistence requires that $\phi_- \leq \phi_0 \leq \phi_+$, it is only possible below the phase transition temperature $T_b = T_c - b\phi_0^2/a$. Thus, for $\phi_0\neq 0$, the order parameter $\delta\phi = (\phi_{\mathrm{max}} - \phi_{\mathrm{min}})/2$ makes a discontinuous jump by $\sqrt{a(T_c-T_b)/(b)}$ at the temperature $T_b$ (first-order phase transition).\footnote{The modern classification of phase transitions based on discontinuities in the order parameter is used here rather than the classical one by Ehrenfest.}
	
	\citet{ThieleFHEKA2019} then proceed to show that, if one considers a \textit{finite} system where interfaces are relevant instead, the Cahn-Hilliard equation still allows us to model the occurrence of demixing in the system. In the finite case, it corresponds to a \textit{bifurcation} (which is the mathematical description of a sudden qualitative change of a system upon a change of a control parameter \citep{Strogatz1994}). The bifurcation diagrams replace the thermodynamic phase diagrams. Bifurcations are commonly studied using a \textit{linear stability analysis}. This method allows us to study the stability of a homogeneous state $\phi_h$ in \cref{cahnh} by considering a small perturbation $\Delta\phi = \phi-\phi_h$, linearizing \cref{cahnh} via a Taylor expansion around $\phi_h$ and solving the resulting linear equation with the ansatz
	\begin{equation}
	\Delta\phi\propto \exp(\lambda(\vec{k}\cdot\vec{r}),
	\end{equation}
	where $\lambda$ is the growth rate, $\vec{k}$ the wavenumber, and $\vec{r}$ the position. The homogeneous state is unstable for $\lambda >0$, since perturbations then grow exponentially. One finds (here in one dimension)
	\begin{equation}
	\lambda(k) = Q(\phi_h)\kappa k^2(k_+^2 - k^2)   
	\end{equation}
	with $k_+^2 = -(a(T-T_c) + 3b\phi_0^2)/\kappa$. In a domain of size $L$, $\lambda$ then becomes positive for 
	\begin{equation}
	\phi_n = \pm \sqrt{\frac{1}{3b}\bigg(a(T_c-T)-\bigg(\kappa\frac{2\pi n}{L}\bigg)^2\bigg)} 
	\end{equation}
	with $n \in \mathbb{N}$. Evidently, we get $\phi_n \to \phi_s$ for $L\to\infty$, i.e., in the thermodynamic limit. The standard description of phase transitions is thus recovered. However, more importantly, one can also describe demixing for the case of finite $L$, and one has obtained a finite-size equivalent of the spinodal. Similarly, a finite-size equivalent of the binodal can also be defined as the concentration value where the inhomogeneous state becomes the global minimum of the free energy. Numerical methods then facilitate the construction of a bifurcation diagram that replaces the phase diagram in the finite system and approaches the phase diagram for increasing $L$. (This paragraph and the previous paragraph closely follow \citet{ThieleFHEKA2019}. See \citet{HollAT2020} for a discussion of this issue for a more complex model.)
	
	What is interesting about this result is that the Cahn-Hilliard equation used for the finite-size system is a simple phenomenological model based on just one order parameter (the field $\phi$), rather than a microscopic statistical description of a many-particle system. Thus, we are clearly in the realm of thermodynamics and not of statistical mechanics.\footnote{One can extend the axioms of thermodynamics by nonequilibrium principles from which theories of the Cahn-Hilliard type can be derived \citep{Doi2013}. In fact, the reciprocal relations of \citet{Onsager1931}, which are crucial in this context, have been referred to as \ZT{fourth law of thermodynamics} \citep{Wendt1974}.} Nevertheless, we are able to describe not only the phase transitions that occur in the limit of an infinite system size, but also the effects that arise from the system actually having a finite size (such as interface contributions). A denial of Callender's proposition 3 should therefore not be taken to imply that thermodynamics is incapable of describing the phenomena we usually call \ZT{phase transitions} in a real system.
	
	In the Cahn-Hilliard equation, the kind of behaviour we call a \ZT{phase transition} in an infinite system is, in a finite system, associated with a bifurcation. Since \ZT{bifurcation} is a perfectly well-defined mathematical concept, this suggests that a re-definition of \ZT{phase transition} applicable to finite systems can be developed based on the idea of a bifurcation. Such a definition then allows us to identify the various real-world effects which we refer to as a \ZT{phase transition} as one natural kind. Alternatively, we can keep the strict standard definition of a \ZT{phase transition}, and only talk about bifurcations if we are interested in finite systems. 
	
	This brings us to the final question of this section, namely what these findings imply for our overall problem. In general, the problem of phase transitions has led philosophers to two sorts of views. Some have denied 3 and thereby interpreted phase transitions as a failure of thermodynamics, while others have denied 4 and thereby interpreted phase transitions as a failure of statistical mechanics (this is then taken to be an argument against reducibility). However, since both thermodynamics and statistical mechanics are very successful theories, seeing either one of them as a failure in the context of phase transitions suggests rather a failure to appreciate what they are capable of. Whether one denies proposition 3 is the merely terminological question of how we define a \ZT{phase transition}. But regardless of whether we call it \ZT{phase transition} or \ZT{bifurcation}, it is clear that thermodynamics is able to model the finite-size phenomenoa associated with phase changes. 
	
	It is therefore an interesting question whether the models used by \citet{ThieleFHEKA2019} are reducible to statistical mechanics. In addition to the simple Cahn-Hilliard equation, they have also considered a phase field crystal (PFC) model. The governing equation of PFC models also has the form \eqref{cahnh}, but the free energy is more complex. The microscopic derivation of PFC models has gained some attention by physicists in recent years \citep{teVrugtW2022}. Therefore, we discuss this issue here based on PFC rather than on phase field models (although also the latter do have a connection to statistical mechanics \citep{MauriB2021}). PFC models are similar to phase field models, but can be used also on atomistic cases to describe crystallization. A review can be found in \citet{EmmerichEtAl2012}. PFC models were first proposed on a phenomenological basis (in a thermodynamic spirit) and then derived from statistical mechanics, namely from a microscopic theory known as \textit{dynamical density functional theory} (DDFT) \citep{teVrugtLW2020} (this derivation is discussed in detail in \citet{ArcherRRS2019} and \citet{teVrugtHKWT2021}).  Thus, we appear to have here a clear case of reduction. However, things are again more complicated than they appear to be at first glance. A study of the solidification of iron by \citet{JaatinenAEA2009}  found that the microscopic predictions of statistical mechanics for the PFC coefficients are not quantitatively accurate. This has to do with the fact that, although PFC models are used to model crystallization, their derivation from statistical mechanics involves certain assumptions (namely small gradients) which are inaccurate for crystallization. On the other hand, one can obtain the PFC parameters by fitting the model to experimental data and thereby use it to describe a wide class of materials. Thus, the reduction of PFC models is an open problem \citep{teVrugtW2022}. On the other hand, the wide applicability of PFC models suggests that they hold for a variety of different microscopic interactions, such that we again have a case of multiple realizability.
	
	
	\section{\label{arrow}The arrow of time}
	The problem of the thermodynamic arrow of time \citep{teVrugt2020}, i.e., the (apparent) incompatibility between the macroscopic laws of thermodynamics (which are irreversible due to the monotonous increase of entropy) and the microscopic laws of classical or quantum mechanics (which are invariant under time reversal) provides a particularly interesting case study for the problem of reduction and fundamentality of thermodynamics. There are (at least) three reasons for this. First, it allows us to find further problems for the standard alleged reduction of thermodynamics. Second, claims that a successful reduction has been achieved here have recently been made in the philosophical literature. Third, it can be used as a possible argument \textit{against} the fundamentality of thermodynamics. This third point will be discussed in \cref{general}.
	
	Let us start with the first point, which can be studied already in the equilibrium case. \citet[p. 545]{Callender2001} describes the reduction of equilibrium thermodynamics to statistical mechanics (which describes many-particle systems using probability distributions) as follows (footnote removed):
	\begin{quotation}
		Thermodynamics assumes something like the following:\\
		Under a given set of environmental conditions (determined externally by
		temperature, pressure, etc.), a system will have approximately constant
		macroscopic properties. A system is in equilibrium just in case it is in such state.\\
		Statistical mechanics ‘translates’ the last claim as:\\
		Thermal equilibrium is a stationary probability distribution.
	\end{quotation}
	He then argues that this view is problematic, since (among other things) it is inappropriate to describe an equilibrium system via a stationary probability distribution such as the microcanonical distribution because a stationary probability distribution is incompatible with the fact that the system has been in nonequilibrium at some point in the past. He thus concludes that we should instead say that
	\begin{quotation}
		Thermal equilibrium corresponds to a special set of microscopic trajectories that leave the macroscopic properties of a system, for a certain observational timescale, approximately constant (and the time scales need not be the same for all macroscopic observables). \citep[p. 547]{Callender2001}
	\end{quotation}
	Interestingly, we are facing here not a problem of thermodynamics, but a problem with its reduction to statistical mechanics. The standard statistical description of an equilibrium system is simply incompatible with the fact that equilibrium systems generally have been nonequilibrium systems at some point in their past.
	
	In general, statistical mechanics will thus have to be based on averaged (\ZT{coarse-grained}) probability distributions, and it is only by using coarse-graining that it is possible to derive macroscopic irreversible dynamics from microscopic reversible theories. This brings us to the second point, which is the reductionist claim recently made by \citet{Robertson2018}. Among the most useful coarse-graining methods is the Mori-Zwanzig formalism \citep{Mori1965,Zwanzig1960,Nakajima1958} (see \citet{Grabert1982} and \citet{teVrugtW2019b} for an introduction), which allows for a systematic derivation of irreversible macroscopic transport equations from reversible microscopic ones. We here briefly sketch this standard derivation following \citet{Zeh1989}, \citet{Wallace2015}, and \citet{teVrugt2021c}. The starting point is the (reversible) Liouville-von Neumann equation for the density operator $\rho$ (describing the system on a microscopic quantum level)
	\begin{equation}
	\dot{\rho}(t)= -\ii L \rho(t), 
	\label{liouville}
	\end{equation}
	with the Liouvillian $L$. We now define a projection operator $P$ that projects $\rho$ onto the \ZT{relevant} part $\bar{\rho}$ (the macroscopic degrees of freedom), i.e., $P\rho=\bar{\rho}$. Moreover, we define an orthogonal projection operator $Q=1-P$ and an irrelevant density $\delta\rho = Q\rho$. This allows the definition of the following \textit{exact} transport equation (see \citet[p. 292]{Wallace2015} and \citet[p. 62]{Zeh1989}):
	\begin{equation}                          
	\dot{\bar{\rho}}(t)=-P\ii L \bar{\rho}(t) + \INT{0}{t}{u}P\ii L e^{-Q\ii L u}Q\ii L \bar{\rho}(t-u) - P\ii L e^{-Q\ii L t}\delta\rho(0).
	\label{exact}
	\end{equation}
	We now note that $P\ii L \bar{\rho}$ vanishes in most cases \citep[p. 62]{Zeh1989} and assume two things, namely
	\begin{enumerate}
		\item that the memory kernel (integrand of the time integral) vanishes very rapidly (Markovian approximation).
		\item that $\delta\rho(0)=0$.
	\end{enumerate}
	This then gives the following \textit{irreversible} transport equation for the macroscopic degrees of freedom (know as \ZT{master equation}) 
	\begin{equation}
	\dot{\bar{\rho}}(t)= \bigg(\INT{0}{\infty}{u}P\ii L e^{-Q\ii L u}Q\ii L\bigg) \bar{\rho}(t).
	\label{higherlevel}
	\end{equation}
	Regarding this derivation, \citet[p. 573]{Robertson2018} says\footnote{In this quote, \ZT{ZZW framework} stands for \ZT{Zwanzig-Zeh-Wallace framework} (Robertson's name for the Mori-Zwanzig formalism), \ZT{SM} stands for \ZT{statistical mechanics}, and \ZT{CM or QM} for \ZT{classical mechanics or quantum mechanics}.}
	\begin{quotation}
		I think that independently of any given account of reduction, this is a case of reduction-in-practice. After all, the ZZW framework allows us to construct the equations of one theory (SM) from another (CM or QM).
	\end{quotation}
	However, things are in fact not so easy for a variety of reasons. 
	
	We start with the second assumption. If we want to derive a theory with an arrow of time (thermodynamics) from a theory without one (statistical mechanics), then we need something to fix the direction of this arrow. By far the most common view is that is to be done by initial conditions, and the assumption $\delta\rho(0)=0$ made here is such an assumption. This is also supported by the fact that seemingly \ZT{anti-thermodynamic behaviour} (heat flowing spontaneously from a cold to a hot spin) can be generated artificially by preparing a system in the \ZT{wrong} initial conditions \citep{MicadeiEtAl2019}. In philosophy, the idea that the thermodynamic asymmetry of time can be explained via an initial condition, namely the low-entropy initial state of the universe, is known as the \ZT{past hypothesis} \citep{Albert2000}. What is a matter of debate is whether the past hypothesis is itself in need of explanation \citep{Callender2016}, which has led some authors to the claim that the past hypothesis itself has the status of a law of nature \citep{Callender2004,Chen2021}. If one adapts this view, then it is clear that we did not derive the laws of thermodynamics from the laws of quantum mechanics, since there was one additional law that is needed (the past hypothesis) which is not among the laws of quantum mechanics.
	
	While this already would make thermodynamics non-reducible to statistical mechanics, it is not obvious that it would make thermodynamics fundamental. One could also argue here that, instead, it is the past hypothesis that is fundamental, and that the past hypothesis is not a part of thermodynamics. Arguments for the latter are (a) the fact that nothing like the past hypothesis appears in the axioms of thermodynamics and (b) that modern formulations of the past hypothesis, backed up by the Mori-Zwanzig derivation \citep{teVrugt2021c}, consider the past hypothesis to be a condition on the initial \textit{microstate} of the universe \citep{Wallace2011}. Moreover, \citet[p. 83]{Dunn2010} has argued that the past hypothesis does not qualify as a (Lewsian) fundamental law since (among other things) it is not a regularity, as a law is required to be. And, finally, it is not even clear that the past hypothesis solves all our problems related to thermodynamic asymmetry. Even if it explains the overall temporal thermodynamic arrow of the universe, it does (perhaps) not explain why it points in the same direction in all subsystems \citep{Winsberg2004}. We can solve all these problems at once by noticing that the minus first law (the assumption that every isolated system spontaneously approaches equilibrium), which incorporates the temporal asymmetry \citep{BrownU2001}, \textit{is} generally considered to be among the laws of thermodynamics, at least in the philosophical literature. The only reason we would need something like the past hypothesis among our fundamental laws is that it generates the thermodynamic asymmetry of time and fixes its direction. It is thus perfectly acceptable to use this asymmetry itself, or more precisely: the minus first law, as a fundamental law instead. Then, the derivation of the irreversible master equation \eqref{higherlevel} in the Mori-Zwanzig formalism sketched above could, very likely, be reduced to a fundamental basis of laws containing quantum mechanics and thermodynamics. Quantum mechanics gives us \cref{exact}, and thermodynamics then justifies the two approximations that lead us from \cref{exact} to \cref{higherlevel}. Thus, for our system of laws---our basis, to use the language introduced in \cref{uff}---to be a \textit{complete} one, we have to add thermodynamics. This makes thermodynamics an element of the CMB and thus a fundamental theory. 
	
	There is a second, more technical problem with considering the derivation of \cref{higherlevel} to be based on nothing but quantum mechanics, which is related to the Markovian approximation. This is a key step, since it generates the irreversibility. However, unless we have a justification for this step within quantum mechanics, then one cannot really say that we have derived irreversible thermodynamics from quantum statistical mechanics. In practice, the Markovian approximation is made because it is known to work, in particular because irreversible transport equations are known to be empirically successful. There is nothing wrong with introducing empirical hypotheses in a derivation, but one should then be careful when claiming that a \ZT{reduction} has been achieved to specify the reducing theory completely. If anything, we have derived statistical mechanics from thermodynamics here, since our knowledge about thermodynamic irreversibility has been used as an input to derive \cref{higherlevel} (a result from statistical mechanics). Understanding the \textit{microscopic} justification of the Markovian approximation is a very challenging problem (see \citet{Toth2020} for a physical and \citet{teVrugt2021c} for a philosophical discussion). Of course, such a justification might be found (research on this problem is ongoing), but we emphasize here that it has to be found if a reduction is to be achieved here. And, notably, some physicists \citep{Drossel2015} have argued that such a justification is in principle impossible.
	
	Interestingly, the Markov approximation also plays a prominent role in \citeauthor{Cartwright1983}'s \citeyearpar{Cartwright1983} discussion of phenomenological and fundamental laws (see in particular her discussion of the master equation in \citet[pp. 113-118]{Cartwright1983}). There, it is used as an example of an approximation that \textit{improves} the underlying theory (here quantum mechanics) in its description of a specific phenomenon, an approximation that is not justified by quantum mechanics itself. While Cartwright's position is certainly more radical than the one defended here, the fact that she uses the Markov approximation as the basis for an argument \textit{against} the view that the fundamental laws (of quantum mechanics) are always true and reproduce the phenomenological laws in specific application contexts should make one question the view that we have an obvious case of reduction here.
	
	To adapt a thought experiment by \citet{Weisskopf1977}\footnote{In its original form, this thought experiment was concerned with whether these physicists could predict the existence of liquids \citep{Weisskopf1977,EvansFD2019}.}, suppose that a group of extremely talented theoretical physicists that has absolutely no empirical knowledge about the world is locked up in a room and presented with the basic laws of quantum mechanics. It has been argued, for example, that these physicists would be able to predict the existence and basic properties of solids \citep{Weisskopf1977}. If this is the case (we will not discuss here whether it is), then it is very plausible that solid state physics can be reduced to quantum mechanics. The interesting question here is whether these physicists would be able to predict that there is such a thing as a thermodynamic arrow of time. Since they would not know anything about the past hypothesis or about the empirical observations that lead to the Markovian assumption, they would presumably \textit{not} be able to predict this.
	
	Note that we do not wish to deny here the tremendous successes of derivations in the Mori-Zwanzig formalism (see, e.g., \citet{teVrugtW2019}, \citet{teVrugtHW2021}, and references therein). For example, they make it possible to obtain microscopic expressions for coefficients appearing as free parameters on the thermodynamic level (such as the viscosity in fluid mechanics \citep{Grabert1982}). Moreover, it is possible to derive novel generalizations of existing theories such as DDFT \citep{WittkowskiLB2012,WittkowskiLB2013}. DDFT, however, also provides a good example for how these derivations work in practice---it was known already from phenomenological work \citep{Evans1979,Munakata1989} and other derivations \citep{MarconiT1999,ArcherE2004} before it was re-derived in the Mori-Zwanzig framework \citep{Yoshimori2005,EspanolL2009} and then finally generalized in this way \citep{WittkowskiLB2012}. Understanding these subtleties is also necessary in order to truly appreciate statistical mechanics, which is not (just) a tool for deriving thermodynamics, but a theory in its own right \citep{Wallace2015}. 

	\section{\label{other}Objections against the fundamentality of thermodynamics}
	\subsection{Microscopic causation}
Often, it is assumed that the most fundamental explanation of an effect is one which is based on microscopic causal processes. In fact, it has been taken to be the basis of the fundamentality of physics that it is causally complete in the sense of providing causal explanations for everything that takes place in the physical world (see \citet{Ney2019}). Does thermodynamics provide causal explanations? \citet[p. 105]{Needham2009} has suggested that it does since, for example, temperature differences \textit{cause} heat to flow from a warmer to a colder object. 

A reservation one might have here is that this would require \ZT{downward causation}---a property of a macroscopic object (here its temperature) determining what its parts do. \citet[p. 33]{Kim1999}, for example, argues that \ZT{higher-level properties can serve as causes in downward causal relations only if they are reducible to lower-level properties}. Typically, it is assumed that fundamental explanations have to be formulated in terms of properties and causal interactions of the microscopic constituents of a system. All macroscopic properties and processes supervene on the microscopic description of the system. However, as \citet[p. 106]{Needham2009} points out, this is based on the assumption that explanations on a lower level are necessarily prior, which itself is in need of explanation. 

In fact, the idea of the priority of microscopic causal explanations appears to be based on a classical picture of the world as consisting of tiny billiard balls that bump into each other according to deterministic laws. The properties of the balls and their bumping are then taken to determine everything that takes place on the macroscopic level. However, the world is actually governed by quantum mechanics. Quantum systems are usually entangled with their environment, and the state of an entangled system can \textit{not} be specified by specifying simply the states of its parts \citep{Maudlin1998}. (Entanglement has, accordingly, been put forward as an argument for monism \citep{Schaffer2010}.) 

The temperature of a many-particle system, while being a statistical property in the classical case, simply depends on the quantum state of the system (as specified by its density operator) in the quantum case. This quantum state is, as we have seen, in most cases not reducible to \ZT{microscopic} properties of the system. Moreover, the heat flow between two systems is determined by their quantum states (as is obvious from the fact that we can reverse the direction of heat flow by manipulating these states \citep{MicadeiEtAl2019}). Thus, from a quantum-mechanical perspective, it is quite natural to argue that thermodynamic processes such as heat flow are determined by properties possessed by the whole rather than by the combined properties of the parts
\footnote{One can also argue \citep{Brenner2018,NaegerS2020,Naeger2020} that the property of being entangled is carried not by the whole, but collectively by its constituents. There are, however, good arguments for considering thermodynamics properties to be possessed by wholes, as discussed in 
\citet{teVrugt2021}.}.

	\subsection{\label{level}Level-dependence of special science laws}
	Avoiding the idea of reduction, \citet{Dunn2010} enumerates a number of conditions intended to show that thermodynamics is not a \ZT{fundamental science}. These are based on \citeauthor{Lewis1983}' \citeyearpar{Lewis1983} account of fundamental laws. In particular, Dunn argues
	\begin{enumerate}
		\item Fundamental laws can only refer to perfectly natural properties. Thermodynamics and the past hypothesis do not satisfy this criterion since they refer to entropy, which is not a perfectly natural property but one related to human observers.
		\item Fundamental laws hold at every level, also between fundamental entities, whereas special science laws hold only at their own level of inquiry. Thermodynamics is solely a macroscopic theory not applicable at the level of a few particles. 
		\item Special science laws are blind to what takes place at the fundamental level. This holds for thermodynamics since it is applies quite independently of what the fundamental laws are. 
	\end{enumerate}
	Objection (1) is based on the fact that entropy, a central quantity of interest in thermodynamics, is closely related to what human observers know about the system (it is often thought of as a measure of \ZT{ignorance}). Dunn bases his discussion of natural laws on the theory of \citet{Lewis1973,Lewis1983}, who understands them as the axioms of a best system, where the \ZT{best system} is the one with the optimal balance of strength and simplicity. These laws have to be formulated only in terms of perfectly natural properties, since otherwise we could make any law simple solely by reformulating it. The axioms of this best system are the fundamental laws \citep{Lewis1983}.\footnote{In a similar vein, \citet[p. 212]{Sider2020} that the \ZT{'fundamentalist' laws of physics} should be expressed via \ZT{fundamental concepts}, where the idea of a \ZT{fundamental concept} is closely related to Lewis' idea of natural properties \citep[p. 2]{Sider2020}.} However, the idea of \ZT{perfectly natural properties} faces several problems. For our purposes, the most important one is that, since the question what is and is not a \ZT{perfectly natural property} is assumed to be a purely metaphysical one, it might be the case that the best candidate for a fundamental theory of physics is not a best system in Lewis' sense if it just happens to be formulated using the \ZT{wrong} properties \citep{Loewer2007,Loewer2021}. Moreover, it is questionable whether \ZT{being a natural property} is in fact a natural property \citep{Thompson2016}, which makes this idea potentially self-defeating.
	
	Moreover, even if we do accept the idea of \ZT{natural properties}, all is not lost here---at least for equilibrium systems, which are the chief realm of thermodynamics. For quantum-mechanical systems in equilibrium, the thermodynamic entropy corresponds to the \textit{von Neumann entropy}, which is defined as
	\begin{equation}
	S = -k_B \Tr(\rho\ln\rho),
	\label{vonneumann}
	\end{equation}
	where $\Tr$ is the quantum-mechanical trace and $\rho$ is the system's density operator. This density operator, in turn, represents an objective description of the state of a quantum-mechanical system and is not necessarily related in any way to what human observers know \citep{Wallace2016}. This stands in contrast to classical statistical mechanics, where $\rho$ represents a probability distribution that \textit{is} generally thought of as representing human credences. However, the world is fundamentally quantum, not classical. Moreover, there are good reasons for understanding the von Neumann entropy \eqref{vonneumann} to correspond to the thermodynamic entropy (see \citet{Chua2021} for a recent defence of this  view). Thus, if $\rho$ is perfectly natural (and there is no reason to assume that it is not), then the (quantum) entropy is too.
	
	Regarding (2), \citet[pp. 86-87]{Dunn2010} says \ZT{fundamental laws hold between entities of fundamental physics, even when we consider these entities singly. [...] These laws of motion, of course, \textit{also} hold between entities of the special sciences}. In fact, (2) actually specifies two requirements: First (2a), fundamental laws hold between the fundamental entities. If we have a criterion for what the fundamental entities are, we can thus identify \ZT{the fundamental laws} as the laws holding for these entities. Second (2b), fundamental laws have to satisfy some sort of generality criterion, according to which they have to hold at every level. Concerning (2b), it should be noted that this criterion is based on a unificationist view of science that not everyone shares. \citet{Cartwright1983,Cartwright1999} is notable in this regard (see \citet{Nager2020b} for an overview). Of course, if we were to adapt such a view, the idea of looking for \ZT{fundamental theory} would need to be revised in general. However, whether (2b) is a reasonable criterion does not really matter here, since (as will be discussed briefly in \cref{general}) thermodynamics does have a pretty wide range of applicability, and thus (2b) will not be the basis for a striking argument against the fundamentality of thermodynamics. What appears to be controversial is whether thermodynamics satisfies (2a).
	
When making a statement regarding the question whether thermodynamics applies to the fundamental entities of physics, we first need to find out what the fundamental entities are. \citet[p. 86]{Dunn2010} does actually mention this problem in a footnote, stating
	
	\begin{quotation}
	  What is a fundamental entity of physics? Following Lewis, we can think of the fundamental entities as those things---whatever they may be---that form the ultimate parts of the Humean mosaic. There is a worry here, that the entities of fundamental physics might not be single particles, but actually massive pluralities of quantum-entangled particles. This possibility is set aside for two reasons. First, we are considering classical correlates of the actual world. Second, it is not clear that [...] this possibility damages any of what is said. 
	\end{quotation}
 
However, this is a bit too quick. While there can be good reasons for studying metaphysical implications of classical systems \citep{teVrugt2021b}, the world is nevertheless governed by quantum mechanics, and if one's argument applies only in a hypothetical world that is not governed by quantum mechanics, then it simply does not apply to our world. And the possibility that this fact damages \ZT{what is said} is quite a realistic one. For if one considers entangled many-particle systems to be objects that are more fundamental than their parts---a position that, although not shared universally (see \citet{Naeger2020}), is not an unreasonable one---then the fundamental entities might be precisely the macroscopic objects that thermodynamics is usually concerned with. In a sense, Dunn's view can thus be subjected to the same criticism that many modern works in mereology have received by \citet{LadymanR2007}, namely that of simply assuming that physics will eventually provide us with a bunch of mereological atoms that everything else is built from like a house is built from bricks, without taking into account that \textit{actual} modern physics does not appear to provide us with any such thing.

Even if we assume that thermodynamics applies only to middle-sized objects, it might still be a fundamental theory by criterion (2b) if we assume that the middle-sized objects are or can be the fundamental ones, a view recently defended by \citet{Inman2017} and \citet{Bernstein2021} (see \citet{Morganti2020}). Let us nevertheless assume, for the sake of the argument, that smaller objects are generally more fundamental. Then, the objection would be that thermodynamics does not apply to small objects. This, however, might have been true for classical \ZT{steam engine thermodynamics}, but not necessarily for modern versions such as quantum thermodynamics \citep{VinjanampathyA2016}. Today, heat flow can also be studied between single atomic nuclei \citep{MicadeiEtAl2019,SchmidtJ2011}. And the black holes discussed earlier should not be forgotten; they are thermodynamic systems yet also very likely to satisfy the criterion of being a fundamental entity.

Finally, (3) is an interesting sort of objection since, for most of this work, we have used the fact that thermodynamics holds regardless of microscopic details as an argument for rather than against the fundamentality of thermodynamics (in particular in our discussion of entropic gravity). The idea that it is characteristic for special sciences to be independent of the fundamental laws certainly stands in a certain tension with the idea that the special sciences should be reducible to these fundamental laws. A problem with Dunn's criterion (3) is, again, that it is helpful for identifying a special science only if we already have an idea of what \ZT{the fundamental level} is. For example, Dunn argues that economics is not a fundamental science because it holds regardless of what is going on at the level of particle physics as long as there is something playing the role of money. However, the reverse is also true---particle physics is blind to what the laws of economics are. If we are (for whatever reason) convinced that economics is a fundamental theory, then criterion (3) would imply that the standard model of particle physics is not. This clearly is a very strange result (particle physics is by all accounts more fundamental than economics), showing that (3) is not a particularly useful criterion. 
	
\subsection{Control theory}
A different (but, as we shall see, related) objection is based on the idea, recently defended by \citet{Wallace2014} and \citet{Myrvold2020b}, that thermodynamics is something like a \ZT{control theory} or \ZT{resource theory}. On this view, which has received support from quantum information theory \citep[p. 145]{Myrvold2021}, thermodynamics is primarily concerned with the effects of certain manipulations of a physical system, and the central concepts of thermodynamics (heat, work, and entropy) are defined relative to a certain class of manipulations. \citet[p. 1220]{Myrvold2020b} argues:
	\begin{quotation}
		A designation of certain variables as manipulable is not something that appears in, or supervenes on, fundamental physics; it must be added. For this reason, $\Theta\Delta^{\mathrm{cs}}$ is not and cannot be a comprehensive or fundamental physical theory.
	\end{quotation}
	This constitutes an argument for why $\Theta\Delta^{\mathrm{cs}}$, which is how \citet{Myrvold2020b} refers to (his understanding of) thermodynamics, is not fundamental. There are (at least) two ways this argument can be met.
	
	First, \citet[p. 1220]{Myrvold2020b} explicitly admits that what he calls $\Theta\Delta^{\mathrm{cs}}$ is a narrower concept than what other authors have understood as \ZT{thermodynamics}. In particular, it does not include the minus first law. \citet[p. 120]{Myrvold2020} advances certain arguments for why the minus first law should not be considered a law of thermodynamics, in particular that it is not concerned with heat and work. There is no room here to discuss this argument, but it should be noted that, as shown in \cref{arrow}, there are good reasons to consider the minus first law to be fundamental. Hence, Myrvold's approach would essentially make thermodynamics a non-fundamental theory simply by removing something fundamental from its axioms. In this case, the non-fundamentality of thermodynamics would result from restricting thermodynamics to what Myrvold calls $\Theta\Delta^{\mathrm{cs}}$. As noted in \cref{fot}, we employ a broader understanding of thermodynamics in this work.
	
	Second, one might question why a resource theory cannot be fundamental. Myrvold appears to take this as given and does not provide an explicit definition of what he takes to be \ZT{fundamental physics}. From the understanding we have proposed in \cref{uff}, it does not follow that a resource theory cannot be fundamental, and if (as is likely, see \cref{entropicgravity,cbht}) thermodynamics can tell us something interesting about quantum gravity, it is quite a move to just take it as axiomatic that theories such as thermodynamics cannot be fundamental. Presumably, the motivation here is similar to the idea discussed in \cref{level} that only perfectly natural properties can feature in fundamental laws---it can be argued that anything that has to do with manipulations is not perfectly natural. However, as has been discussed in \cref{level}, we do not subscribe to the naturalness criterion in this paper. Moreover, \citet[p. 1222]{Myrvold2020b} strongly objects against the view that the $\Theta\Delta^{\mathrm{cs}}$ is subjective, in part because the variables treated as manipulable do not have to be manipulable by humans (he cites applications to volcanoes as an example). We agree to this argument, and therefore maintain that thermodynamics is a fundamental theory. 
	
	\subsection{\label{general}Generality}
	Another criterion for \ZT{fundamentality} that is behind several recent proposals (such as explanatory maximality) is that a fundamental theory should be one that has a wide range of applicability compared to non-fundamental theories. We should therefore also take a brief look at how thermodynamics fares in this regard. First, it is sometimes argued that thermodynamics applies only to macroscopic systems. As discussed in \citet{teVrugt2021} and in \cref{level}, this is a prejudice that is based on not considering modern approaches to the thermodynamics of small systems, such as quantum thermodynamics \citep{VinjanampathyA2016} or stochastic thermodynamics \citep{Seifert2012}. Second, most of thermodynamics applies only to equilibrium systems. However, there are ways to apply it also in certain regions away from equilibrium (such as nonequilibrium thermodynamics), and the minus first law is a general statement of thermodynamics applying to nonequilibrium systems (stating that they will approach equilibrium). In particular, stochastic thermodynamics is applicable to small non-equilibrium systems \citep{Seifert2012}---applications to active matter \citep{Speck2016} are a good example. As mentioned in \cref{fot}, we consider all these more recent extensions to be a part of thermodynamics. One can adopt a more restricted definition, but again, it does not seem reasonable to us to defend the view that thermodynamics is not fundamental on the basis of excluding a large part of what, in modern physics, is considered an important part of thermodynamics.
	
	The question is then whether there are isolated systems that do not approach equilibrium or even move away from it. A potentially problematic candidate would be classical self-gravitating systems, where temperature gradients sometimes increase due to a feedback loop arising from negative heat capacity, an effect known as \ZT{gravothermal catastrophe} \citep{LyndenBellW1968}. \citet{Robertson2019} has argued that self-gravitating systems can be described by statistical mechanics but not by thermodynamics, and has therefore expressed scepticism regarding the fundamentality of thermodynamics. However, a real sufficiently large star will eventally collapse to a black hole, which \textit{is} an equilibrium state of a self-gravitating system \citep{Wallace2010}. Moreover, the view that self-gravitating systems do not exhibit thermodynamic behaviour might be due to too narrow a conception of what \ZT{thermodynamics} is \citep{Callender2011}. Another objection would consider systems moving away from equilibrium due to special initial correlations. (Performing the derivation described in \cref{arrow} with a different initial condition than the one found to be satisfied in physical systems can produce a decreasing entropy \citep[p. 67]{Zeh1989}.) Whether setups such as the spin echo experiment \citep{Hahn1950}, where a disordered system of spins spontaneously evolves into an ordered one, count as an exception \citep{RidderbosR1998} is controversial. But even if one takes this view, such exceptions are rare and arise only in highly controlled laboratory environments specifically designed to create them. Regardless of one's view on the spin echo experimental, it is clear that thermodynamics is applicable much more broadly than almost any other theory. 
	
	It should be noted here that even quantum field theory and general relativity---which, among all presently used and empirically confirmed scientific theories, are the ones that are most likely to be fundamental---cannot hold everywhere since they are incompatible. (One could therefore reserve the word \ZT{fundamental} solely for a yet to be derived exact theory of quantum gravity. But, if we use a definition of \ZT{fundamental} that will only be satisfied by a scientific theory that is centuries away or will never be reached, then the word \ZT{fundamental} becomes more of less useless for talking about current science \citep{Ney2019}.) In contrast, thermodynamics applies both to systems governed by quantum field theory and in cosmological contexts where general relativity is relevant, and is often taken to be a useful tool in approaching quantum gravity (as has been discussed earlier). Thus, if we count the spin echo experiment as a violation of thermodynamics (which, as just noted, is not something that everyone does), and we discard thermodynamics as a candidate for a fundamental theory because of this exception, then we also have to discard quantum field theory and general relativity because they will also break down somewhere. Alternatively (and more reasonably), we can weaken the requirement of being an exceptionless theory. In this case, however, we have every reason to think of thermodynamics as a fundamental theory.

	\section{\label{conclusion}Conclusion}
	It is widely believed among philosophers of physics (and philosophers of science in general) that the relation of thermodynamics to statistical mechanics is a clear case (perhaps the clearest case) of intertheoretical reduction. The main purpose of this paper is to pour a little cold water on this claim. In general, the alleged reduction of thermodynamics is heavily affected by the multiple realizability problem, an issue for which entropic gravity and black hole thermodynamics provide a much more striking example than the commonly discussed scenarios. Moreover, apparent explicit reductions such as the derivation of phase field crystal models or the derivation of irreversibility in the Mori-Zwanzig formalism, do, at a closer look, tend to involve auxiliary assumptions that are motivated primarily by the anticipated reduction. Finally, we have sketched how recent results in phase field modelling might allow for a thermodynamic description of phase transitions even beyond the thermodynamic limit.

	\begin{acknowledgements}
		We thank Lodin Ellingsen, Tobias Frohoff-H\"ulsmann, Tore Haug-Warberg, Ulrich Krohs, Paul M. N\"ager, Lucia Oliveri, Peter Rohs, Oliver Robert Scholz, Ansgar Seide, and Raphael Wittkowski for helpful discussions. M.t.V. thanks the Studienstiftung des Deutschen Volkes for financial support. G.J.S. thanks for support by the Deutsche Forschungsgemeinschaft [EXC 2023, Project-ID: 39062112].
	\end{acknowledgements}


\begin{thebibliography}{148}
\providecommand{\natexlab}[1]{#1}
\providecommand{\url}[1]{{#1}}
\providecommand{\urlprefix}{URL }
\expandafter\ifx\csname urlstyle\endcsname\relax
  \providecommand{\doi}[1]{DOI~\discretionary{}{}{}#1}\else
  \providecommand{\doi}{DOI~\discretionary{}{}{}\begingroup
  \urlstyle{rm}\Url}\fi
\providecommand{\eprint}[2][]{\url{#2}}

\bibitem[{Abbott and others {(LIGO Scientific Collaboration and Virgo
  Collaboration)}(2016)}]{Abbott2016}
Abbott BP, others {(LIGO Scientific Collaboration and Virgo Collaboration)}
  (2016) Observation of gravitational waves from a binary black hole merger.
  \emph{Physical Review Letters} 116(6):061102

\bibitem[{Albert(2000)}]{Albert2000}
Albert DZ (2000) \emph{Time and Chance}. Harvard University Press, Cambridge
  (Massachusetts)

\bibitem[{Allen and Cahn(1976)}]{AllenC1976}
Allen SM, Cahn JW (1976) On tricritical points resulting from the intersection
  of lines of higher-order transitions with spinodals. \emph{Scripta
  Metallurgica} 10(5):451--454

\bibitem[{Archer and Evans(2004)}]{ArcherE2004}
Archer AJ, Evans R (2004) Dynamical density functional theory and its
  application to spinodal decomposition. \emph{Journal of Chemical Physics}
  121(9):4246--4254

\bibitem[{Archer et~al.(2019)Archer, Ratliff, Rucklidge, and
  Subramanian}]{ArcherRRS2019}
Archer AJ, Ratliff DJ, Rucklidge AM, Subramanian P (2019) Deriving phase field
  crystal theory from dynamical density functional theory: consequences of the
  approximations. \emph{Physical Review E} 100(2):022140

\bibitem[{Bardeen et~al.(1973)Bardeen, Carter, and Hawking}]{BardeenCH1973}
Bardeen JM, Carter B, Hawking SW (1973) The four laws of black hole mechanics.
  \emph{Communications in Mathematical Physics} 31(2):161--170

\bibitem[{Batterman(2000)}]{Batterman2000}
Batterman RW (2000) Multiple realizability and universality. \emph{British
  Journal for the Philosophy of Science} 51(1):115--145

\bibitem[{Bekenstein(1972)}]{Bekenstein1972}
Bekenstein JD (1972) Black holes and the second law. \emph{Lettere al Nuovo
  Cimento} 4:737--740

\bibitem[{Bekenstein(1973)}]{Bekenstein1973}
Bekenstein JD (1973) Black holes and entropy. \emph{Physical Review D}
  7:2333--2346

\bibitem[{Bennett(2017)}]{Bennett2017}
Bennett K (2017) \emph{Making things up}. Oxford University Press, Oxford

\bibitem[{Bernstein(2021)}]{Bernstein2021}
Bernstein S (2021) Could a middle level be the most fundamental?
  \emph{Philosophical Studies} 178(4):1065--1078

\bibitem[{Bickle(1998)}]{Bickle1998}
Bickle J (1998) \emph{Psychoneural reduction: The new wave}. MIT Press,
  Cambridge, Massachusetts

\bibitem[{Bickle(2020)}]{Bickle2020}
Bickle J (2020) \emph{{Multiple Realizability}}. In: Zalta EN (ed) The
  {Stanford} Encyclopedia of Philosophy, {S}ummer 2020 edn, Metaphysics
  Research Lab, Stanford University

\bibitem[{Bliss and Trogdon(2021)}]{BlissT2021}
Bliss R, Trogdon K (2021) \emph{{Metaphysical Grounding}}. In: Zalta EN (ed)
  The {Stanford} Encyclopedia of Philosophy, {W}inter 2021 edn, Metaphysics
  Research Lab, Stanford University

\bibitem[{Brenner(2018)}]{Brenner2018}
Brenner A (2018) Science and the special composition question. \emph{Synthese}
  195(2):657--678

\bibitem[{Brown(2005)}]{Brown2005}
Brown HR (2005) \emph{Physical Relativity: Space-time Structure from a
  Dynamical Perspective}. Clarendon Press, Oxford

\bibitem[{Brown and Uffink(2001)}]{BrownU2001}
Brown HR, Uffink J (2001) The origins of time-asymmetry in thermodynamics: The
  minus first law. \emph{Studies in History and Philosophy of Modern Physics}
  32(4):525 -- 538

\bibitem[{Cahn(1965)}]{Cahn1965}
Cahn JW (1965) Phase separation by spinodal decomposition in isotropic systems.
  \emph{Journal of Chemical Physics} 42(1):93--99

\bibitem[{Cahn and Hilliard(1958)}]{CahnH1958}
Cahn JW, Hilliard JE (1958) Free energy of a nonuniform system. i. interfacial
  free energy. \emph{Journal of Chemical Physics} 28(2):258--267

\bibitem[{Callender(2001)}]{Callender2001}
Callender C (2001) Taking thermodynamics too seriously. \emph{Studies in
  History and Philosophy of Modern Physics} 32(4):539--553

\bibitem[{Callender(2004)}]{Callender2004}
Callender C (2004) Measures, explanations and the past: Should 'special'
  initial conditions be explained? \emph{British Journal for the Philosophy of
  Science} 55(2):195--217

\bibitem[{Callender(2011)}]{Callender2011}
Callender C (2011) Hot and heavy matters in the foundations of statistical
  mechanics. \emph{Foundations of Physics} 41(6):960--981

\bibitem[{Callender(2016)}]{Callender2016}
Callender C (2016) \emph{Thermodynamic Asymmetry in Time}. In: Zalta EN (ed)
  The Stanford Encyclopedia of Philosophy, winter 2016 edn, The Stanford
  Encyclopedia of Philosophy, Metaphysics Research Lab, Stanford University

\bibitem[{Cartwright(1983)}]{Cartwright1983}
Cartwright N (1983) \emph{How the Laws of Physics Lie}. Oxford University
  Press, Oxford

\bibitem[{Cartwright(1999)}]{Cartwright1999}
Cartwright N (1999) \emph{The Dappled World: A Study of the Boundaries of
  Science}. Cambridge University Press, Cambridge

\bibitem[{Chen(2021)}]{Chen2021}
Chen EK (2021) Quantum mechanics in a time-asymmetric universe: On the nature
  of the initial quantum state. \emph{British Journal for the Philosophy of
  Science} 72(4):1155--1183

\bibitem[{Chua(2021)}]{Chua2021}
Chua EYS (2021) Does von {N}eumann entropy correspond to thermodynamic entropy?
  \emph{Philosophy of Science} 88(1):145--168

\bibitem[{Clapp(2001)}]{Clapp2001}
Clapp L (2001) Disjunctive properties: Multiple realizations. \emph{Journal of
  Philosophy} 98(3):111--136

\bibitem[{Crowther(2020)}]{Crowther2020}
Crowther K (2020) What is the point of reduction in science? \emph{Erkenntnis}
  85(6):1437--1460

\bibitem[{Curiel(2014)}]{Curiel2014}
Curiel E (2014) Classical black holes are hot. \emph{arXiv:14083691}

\bibitem[{{de Groot} and Mazur(1962)}]{deGrootM1962}
{de Groot} SR, Mazur P (1962) \emph{Non-equilibrium Thermodynamics}.
  North-Holland, Amsterdam

\bibitem[{Dieks et~al.(2015)Dieks, {van Dongen}, and {de Haro}}]{DieksvDdH2015}
Dieks D, {van Dongen} J, {de Haro} S (2015) Emergence in holographic scenarios
  for gravity. \emph{Studies in History and Philosophy of Modern Physics}
  52:203--216

\bibitem[{Doi(2013)}]{Doi2013}
Doi M (2013) \emph{Soft Matter Physics}. Oxford University Press, Oxford

\bibitem[{Dougherty and Callender(2016)}]{DoughertyC2016}
Dougherty J, Callender C (2016) Black hole thermodynamics: More than an
  analogy? \emph{preprint} Available at
  \texttt{\seqsplit{http://philsci-archive.pitt.edu/13195/}}

\bibitem[{Drossel(2015)}]{Drossel2015}
Drossel B (2015) \emph{On the Relation Between the Second Law of Thermodynamics
  and Classical and Quantum Mechanics}. In: Falkenburg B, Morrison M (eds) Why
  More Is Different: Philosophical Issues in Condensed Matter Physics and
  Complex Systems, Springer, Berlin, pp 41--54

\bibitem[{Dunn(2011)}]{Dunn2010}
Dunn J (2011) Fried eggs, thermodynamics, and the special sciences.
  \emph{British Journal for the Philosophy of Science} 62(1):71--98

\bibitem[{Eddington(1935)}]{Eddington1935}
Eddington A (1935) \emph{The Nature of the Physical World}. Everyman’s
  Library, J. M. Dent, London

\bibitem[{Einstein(1970)}]{Einstein1970}
Einstein A (1970) \emph{Autobiographical Notes}. In: Schilpp PA (ed) Albert
  Einstein: Philosopher-Scientist, Vol. 2, Cambridge University Press,
  Cambridge

\bibitem[{Emmerich(2008)}]{Emmerich2008}
Emmerich H (2008) Advances of and by phase-field modelling in condensed-matter
  physics. \emph{Advances in Physics} 57(1):1--87

\bibitem[{Emmerich et~al.(2012)Emmerich, L{\"o}wen, Wittkowski, Gruhn,
  T{\'o}th, Tegze, and Gr{\'a}n{\'a}sy}]{EmmerichEtAl2012}
Emmerich H, L{\"o}wen H, Wittkowski R, Gruhn T, T{\'o}th GI, Tegze G,
  Gr{\'a}n{\'a}sy L (2012) Phase-field-crystal models for condensed matter
  dynamics on atomic length and diffusive time scales: an overview.
  \emph{Advances in Physics} 61(6):665--743

\bibitem[{Espa{\~n}ol and L\"owen(2009)}]{EspanolL2009}
Espa{\~n}ol P, L\"owen H (2009) Derivation of dynamical density functional
  theory using the projection operator technique. \emph{Journal of Chemical
  Physics} 131(24):244101

\bibitem[{Evans(1979)}]{Evans1979}
Evans R (1979) The nature of the liquid-vapour interface and other topics in
  the statistical mechanics of non-uniform, classical fluids. \emph{Advances in
  Physics} 28:143--200

\bibitem[{Evans et~al.(2019)Evans, Frenkel, and Dijkstra}]{EvansFD2019}
Evans R, Frenkel D, Dijkstra M (2019) From simple liquids to colloids and soft
  matter. \emph{Physics Today} 72(38.10):1063

\bibitem[{Fodor(1974)}]{Fodor1974}
Fodor JA (1974) Special sciences (or: The disunity of science as a working
  hypothesis). \emph{Synthese} 28:97--115

\bibitem[{Grabert(1982)}]{Grabert1982}
Grabert H (1982) \emph{Projection Operator Techniques in Nonequilibrium
  Statistical Mechanics}, Springer Tracts in Modern Physics, vol~95, 1st edn.
  Springer-Verlag, Berlin

\bibitem[{Hahn(1950)}]{Hahn1950}
Hahn EL (1950) Spin echoes. \emph{Physical Review} 80(4):580

\bibitem[{Holl et~al.(2020)Holl, Archer, and Thiele}]{HollAT2020}
Holl MP, Archer AJ, Thiele U (2020) Efficient calculation of phase coexistence
  and phase diagrams: application to a binary phase-field crystal model.
  \emph{Journal of Physics: Condensed Matter} 33(11):115401

\bibitem[{Inman(2017)}]{Inman2017}
Inman RD (2017) \emph{Substance and the fundamentality of the familiar: a
  neo-{A}ristotelian mereology}. Routledge, New York

\bibitem[{Jaatinen et~al.(2009)Jaatinen, Achim, Elder, and
  Ala-Nissila}]{JaatinenAEA2009}
Jaatinen A, Achim CV, Elder KR, Ala-Nissila T (2009) Thermodynamics of bcc
  metals in phase-field-crystal models. \emph{Physical Review E} 80(3):031602

\bibitem[{Jaeger(1998)}]{Jaeger1998}
Jaeger G (1998) The {E}hrenfest classification of phase transitions:
  introduction and evolution. \emph{Archive for History of Exact Sciences}
  53(1):51--81

\bibitem[{Kadanoff(2009)}]{Kadanoff2009}
Kadanoff LP (2009) More is the same; phase transitions and mean field theories.
  \emph{Journal of Statistical Physics} 137(5):777--797

\bibitem[{Kim(1989)}]{Kim1989}
Kim J (1989) The myth of nonreductive materialism. \emph{Proceedings and
  Addresses of the American Philosophical Association} 63(3):31--47

\bibitem[{Kim(1992)}]{Kim1992}
Kim J (1992) Multiple realization and the metaphysics of reduction.
  \emph{Philosophy and Phenomenological Research} 52(1):1--26

\bibitem[{Kim(1999)}]{Kim1999}
Kim J (1999) Making sense of emergence. \emph{Philosophical Studies}
  95(1/2):3--36

\bibitem[{Ladyman and Ross(2007)}]{LadymanR2007}
Ladyman J, Ross D (2007) \emph{Every thing must go: Metaphysics naturalized}.
  Oxford University Press, Oxford

\bibitem[{Lewis(1973)}]{Lewis1973}
Lewis D (1973) \emph{Counterfactuals}. Harvard University Press, Cambridge

\bibitem[{Lewis(1983)}]{Lewis1983}
Lewis D (1983) New work for a theory of universals. \emph{Australasian Journal
  of Philosophy} 61(4):343--377

\bibitem[{Lewis(1986)}]{Lewis1986}
Lewis D (1986) \emph{On the plurality of worlds}. Blackwell, Oxford

\bibitem[{Liu(1999)}]{Liu1999}
Liu C (1999) Explaining the emergence of cooperative phenomena.
  \emph{Philosophy of Science} 66:S92--S106

\bibitem[{Loewer(2007)}]{Loewer2007}
Loewer B (2007) Laws and natural properties. \emph{Philosophical Topics}
  35(1/2):313--328

\bibitem[{Loewer(2021)}]{Loewer2021}
Loewer B (2021) The package deal account of laws and properties (pda).
  \emph{Synthese} 199(1):1065--1089

\bibitem[{Lynden-Bell and Wood(1968)}]{LyndenBellW1968}
Lynden-Bell D, Wood R (1968) The gravo-thermal catastrophe in isothermal
  spheres and the onset of red-giant structure for stellar systems.
  \emph{Monthly Notices of the Royal Astronomical Society} 138(4):495--525

\bibitem[{Mandal et~al.(2019)Mandal, Liebchen, and L{\"o}wen}]{MandalLL2019}
Mandal S, Liebchen B, L{\"o}wen H (2019) Motility-induced temperature
  difference in coexisting phases. \emph{Physical Review Letters}
  123(22):228001

\bibitem[{{Marini Bettolo Marconi} and Tarazona(1999)}]{MarconiT1999}
{Marini Bettolo Marconi} U, Tarazona P (1999) Dynamic density functional theory
  of fluids. \emph{Journal of Chemical Physics} 110(16):8032--8044

\bibitem[{Marras(2002)}]{Marras2002}
Marras A (2002) Kim on reduction. \emph{Erkenntnis} 57(2):231--257

\bibitem[{Maudlin(1998)}]{Maudlin1998}
Maudlin T (1998) \emph{Part and Whole in Quantum Mechanics}. In: Castellani E
  (ed) Interpreting Bodies, Princeton University Press, Princeton, pp 46--60

\bibitem[{Mauri and Bertei(2021)}]{MauriB2021}
Mauri R, Bertei A (2021) Non-local phase field revisited. \emph{Journal of
  Statistical Mechanics: Theory and Experiment} 2021(6):063212

\bibitem[{Menon and Callender(2013)}]{MenonC2013}
Menon T, Callender C (2013) \emph{Turn and Face The Strange … Ch-Ch-Changes:
  Philosophical Questions Raised by Phase Transitions}. In: Batterman R (ed)
  The {O}xford Handbook of Philosophy of Physics, Oxford University Press,
  Oxford, pp 190--223

\bibitem[{Micadei et~al.(2019)Micadei, Peterson, Souza, Sarthour, Oliveira,
  Landi, Batalh{\~a}o, Serra, and Lutz}]{MicadeiEtAl2019}
Micadei K, Peterson JPS, Souza AM, Sarthour RS, Oliveira IS, Landi GT,
  Batalh{\~a}o TB, Serra RM, Lutz E (2019) Reversing the direction of heat flow
  using quantum correlations. \emph{Nature Communications} 10(1):2456

\bibitem[{{MICRESS}(2021)}]{Micress}
{MICRESS} (2021) {The MICRostructure Evolution Simulation Software}.
  \texttt{\seqsplit{www.micress.de}} (accessed Feb 20th 2021)

\bibitem[{Milgrom(1983)}]{Milgrom1983}
Milgrom M (1983) A modification of the {N}ewtonian dynamics as a possible
  alternative to the hidden mass hypothesis. \emph{Astrophysical Journal}
  270:365--370

\bibitem[{Morganti(2020{\natexlab{a}})}]{Morganti2020}
Morganti M (2020{\natexlab{a}}) Fundamentality in metaphysics and the
  philosophy of physics. {P}art {I}: {M}etaphysics. \emph{Philosophy Compass}
  15(7):e12690

\bibitem[{Morganti(2020{\natexlab{b}})}]{Morganti2020b}
Morganti M (2020{\natexlab{b}}) Fundamentality in metaphysics and the
  philosophy of physics. {P}art {II}: {T}he philosophy of physics.
  \emph{Philosophy Compass} 15(10):e12703

\bibitem[{Mori(1965)}]{Mori1965}
Mori H (1965) Transport, collective motion, and {B}rownian motion.
  \emph{Progress of Theoretical Physics} 33(3):423--455

\bibitem[{Munakata(1989)}]{Munakata1989}
Munakata T (1989) A dynamical extension of the density functional theory.
  \emph{Journal of the Physical Society of Japan} 58(7):2434--2438

\bibitem[{Myrvold(2020{\natexlab{a}})}]{Myrvold2020}
Myrvold WC (2020{\natexlab{a}}) \emph{Explaining Thermodynamics: What Remains
  to be Done?} In: Allori V (ed) Statistical Mechanics and Scientific
  Explanation, World Scientific, Singapore, pp 113--143

\bibitem[{Myrvold(2020{\natexlab{b}})}]{Myrvold2020b}
Myrvold WC (2020{\natexlab{b}}) The science of {$\Theta\Delta^{\mathrm{cs}}$}.
  \emph{Foundations of Physics} 50(10):1219--1251

\bibitem[{Myrvold(2021)}]{Myrvold2021}
Myrvold WC (2021) \emph{Beyond chance and credence: A theory of hybrid
  probabilities}. Oxford University Press, Oxford

\bibitem[{Nagel(1961)}]{Nagel1961}
Nagel E (1961) \emph{The Structure of Science}. Harcourt, Brace \& World, New
  York

\bibitem[{N{\"a}ger(2020)}]{Nager2020b}
N{\"a}ger PM (2020) \emph{Was ist ein {N}aturgesetz? {N}ancy {C}artwright}. In:
  M\"uller-Salo J (ed) Analytische Philosophie: Eine Einf{\"u}hrung in 16
  Fragen und Antworten, UTB, Paderborn, pp 57--66

\bibitem[{N\"ager(2021)}]{Naeger2020}
N\"ager PM (2021) The mereological problem of entanglement. \emph{preprint}
  Available at \texttt{\seqsplit{https://philarchive.org/rec/NGETMP}}

\bibitem[{N\"ager and Strobach(2021)}]{NaegerS2020}
N\"ager PM, Strobach N (2021) A taxonomy for the mereology of entangled quantum
  systems. \emph{preprint} Available at
  \texttt{\seqsplit{https://philarchive.org/rec/NGEATF}}

\bibitem[{Nakajima(1958)}]{Nakajima1958}
Nakajima S (1958) On quantum theory of transport phenomena: steady diffusion.
  \emph{Progress of Theoretical Physics} 20(6):948--959

\bibitem[{Needham(2009)}]{Needham2009}
Needham P (2009) Reduction and emergence: a critique of {K}im.
  \emph{Philosophical Studies} 146(1):93--116

\bibitem[{Needham(2010)}]{Needham2010}
Needham P (2010) Nagel's analysis of reduction: Comments in defense as well as
  critique. \emph{Studies in History and Philosophy of Modern Physics}
  41(2):163--170

\bibitem[{Needham(2013)}]{Needham2013}
Needham P (2013) Process and change: From a thermodynamic perspective.
  \emph{British Journal for the Philosophy of Science} 64(2):395--422

\bibitem[{Needham(2018)}]{Needham2018}
Needham P (2018) \emph{Macroscopic metaphysics: Middle-sized objects and
  longish processes}. Springer, Cham

\bibitem[{Ney(2019)}]{Ney2019}
Ney A (2019) \emph{The politics of fundamentality}. In: Aguirre A, Foster B,
  Merali Z (eds) What is Fundamental?, Springer, Cham, pp 27--36

\bibitem[{North(2011)}]{North2011}
North J (2011) \emph{Time in thermodynamics}. In: Callender C (ed) The Oxford
  Handbook of Philosophy of Time, Oxford University Press, pp 312--350

\bibitem[{Onsager(1931)}]{Onsager1931}
Onsager L (1931) Reciprocal relations in irreversible processes. i.
  \emph{Physical Review} 37:405--426

\bibitem[{Oppenheim and Putnam(1958)}]{OppenheimP1958}
Oppenheim P, Putnam H (1958) \emph{The unity of science as a working
  hypothesis}. In: Feigl H, Scriven M, Maxwell G (eds) Minnesota Studies in the
  Philosophy of Science, vol. 2, Minnesota University Press, Minneapolis

\bibitem[{Place(1956)}]{Place1956}
Place UT (1956) Is consciousness a brain process? \emph{British Journal of
  Psychology} 47(1):44--50

\bibitem[{Prigogine(1997)}]{PrigogineS1997}
Prigogine I (1997) \emph{The end of certainty}. The Free Press, New York

\bibitem[{Provatas and Elder(2011)}]{ProvatasE2011}
Provatas N, Elder K (2011) \emph{Phase-field methods in materials science and
  engineering}. Wiley VCH, Weinheim

\bibitem[{Putnam(1967)}]{Putnam1967}
Putnam H (1967) \emph{Psychological predicates}. In: Capitan WH, Merrill DD
  (eds) Art, Mind, and Religion, University of Pittsburgh Press, Pittsburgh, pp
  37--48

\bibitem[{Ridderbos and Redhead(1998)}]{RidderbosR1998}
Ridderbos TM, Redhead MLG (1998) The spin-echo experiments and the second law
  of thermodynamics. \emph{Foundations of Physics} 28(8):1237--1270

\bibitem[{Robertson(2019)}]{Robertson2019}
Robertson K (2019) Stars and steam engines: To what extent do thermodynamics
  and statistical mechanics apply to self-gravitating systems? \emph{Synthese}
  196(5):1783--1808

\bibitem[{Robertson(2020)}]{Robertson2018}
Robertson K (2020) Asymmetry, abstraction, and autonomy: Justifying
  coarse-graining in statistical mechanics. \emph{British Journal for the
  Philosophy of Science} 71(2):547--579

\bibitem[{Schaffer(2010)}]{Schaffer2010}
Schaffer J (2010) Monism: The priority of the whole. \emph{Philosophical
  Review} 119(1):31--76

\bibitem[{Schmidt and Jurado(2011)}]{SchmidtJ2011}
Schmidt KH, Jurado B (2011) Thermodynamics of nuclei in thermal contact.
  \emph{Physical Review C} 83(1):014607

\bibitem[{Schmitz(2003)}]{Schmitz2003}
Schmitz GJ (2003) \emph{Thermodynamics of Diffuse Interfaces}. In: Emmerich H,
  Nestler B, Schreckenberg M (eds) Interface and Transport Dynamics, Springer,
  Berlin Heidelberg, pp 47--64

\bibitem[{Schmitz(2017)}]{Schmitz2017}
Schmitz GJ (2017) A combined entropy/phase-field approach to gravity.
  \emph{Entropy} 19(4):151

\bibitem[{Schmitz(2018)}]{Schmitz2018}
Schmitz GJ (2018) Entropy of geometric objects. \emph{Entropy} 20(6):453

\bibitem[{Schmitz(2020)}]{Schmitz2020}
Schmitz GJ (2020) Quantitative mereology: An essay to align physics laws with a
  philosophical concept. \emph{Physics Essays} 33(4):479--488

\bibitem[{Schmitz(2022)}]{Schmitz2022}
Schmitz GJ (2022) A phase-field perspective on mereotopology.
  \emph{AppliedMath} 2(1):54--104

\bibitem[{Schmitz et~al.(2022)Schmitz, {te Vrugt}, Haug-Warberg, Ellingsen,
  Needham, and Wittkowski}]{SchmitztVHWENW2022}
Schmitz GJ, {te Vrugt} M, Haug-Warberg T, Ellingsen L, Needham P, Wittkowski R
  (2022) Thermodynamics of an empty box. \emph{in preparation}

\bibitem[{Scholz(2018)}]{Scholz2019}
Scholz OR (2018) \emph{{Induktive Metaphysik – Ein vergessenes Kapitel der
  Metaphysikgeschichte}}. In: Hommen D, S\"olch D (eds) Philosophische Sprache
  zwischen Tradition und Innovation (Festschrift f\"ur Christoph Kann), Peter
  Lang, Berlin, pp 267--289

\bibitem[{Seifert(2012)}]{Seifert2012}
Seifert U (2012) Stochastic thermodynamics, fluctuation theorems and molecular
  machines. \emph{Reports on Progress in Physics} 75(12):126001

\bibitem[{Sider(2020)}]{Sider2020}
Sider T (2020) \emph{The tools of metaphysics and the metaphysics of science}.
  Oxford University Press, Oxford

\bibitem[{Smart(1959)}]{Smart1959}
Smart JJC (1959) Sensations and brain processes. \emph{Philosophical Review}
  68(2):141--156

\bibitem[{Speck(2016)}]{Speck2016}
Speck T (2016) Stochastic thermodynamics for active matter. \emph{Europhysics
  Letters} 114(3):30006

\bibitem[{Steinbach(2009)}]{Steinbach2009}
Steinbach I (2009) Phase-field models in materials science. \emph{Modelling and
  Simulation in Materials Science and Engineering} 17(7):073001

\bibitem[{Strogatz(1994)}]{Strogatz1994}
Strogatz SH (1994) \emph{Nonlinear Dynamics and Chaos}. Perseus, New York

\bibitem[{Tadmor et~al.(2012)Tadmor, Miller, and Elliott}]{TadmorME2012}
Tadmor EB, Miller RE, Elliott RS (2012) \emph{Continuum mechanics and
  thermodynamics: from fundamental concepts to governing equations}. Cambridge
  University Press, Cambridge

\bibitem[{Tahko(2018)}]{Tahko2018}
Tahko TE (2018) \emph{{Fundamentality}}. In: Zalta EN (ed) The {Stanford}
  Encyclopedia of Philosophy, {F}all 2018 edn, Metaphysics Research Lab,
  Stanford University

\bibitem[{{te Vrugt}(2021{\natexlab{a}})}]{teVrugt2020}
{te Vrugt} M (2021{\natexlab{a}}) The five problems of irreversibility.
  \emph{Studies in History and Philosophy of Science} 87:136--146

\bibitem[{{te Vrugt}(2021{\natexlab{b}})}]{teVrugt2021}
{te Vrugt} M (2021{\natexlab{b}}) The mereology of thermodynamic equilibrium.
  \emph{Synthese} 199:12891--12921

\bibitem[{{te Vrugt}(2021{\natexlab{c}})}]{teVrugt2021c}
{te Vrugt} M (2021{\natexlab{c}}) Understanding probability and irreversibility
  in the {M}ori-{Z}wanzig projection operator formalism. \emph{arXiv:211204067}

\bibitem[{{te Vrugt}(forthcoming)}]{teVrugt2021b}
{te Vrugt} M (forthcoming) How to distinguish between indistinguishable
  particles. \emph{British Journal for the Philosophy of Science} Available at
  \texttt{\seqsplit{https://doi.org/10.1086/718495}}

\bibitem[{{te Vrugt} and Wittkowski(2019)}]{teVrugtW2019}
{te Vrugt} M, Wittkowski R (2019) {Mori-Zwanzig projection operator formalism
  for far-from-equilibrium systems with time-dependent Hamiltonians}.
  \emph{Physical Review E} 99:062118

\bibitem[{{te Vrugt} and Wittkowski(2020)}]{teVrugtW2019b}
{te Vrugt} M, Wittkowski R (2020) Projection operators in statistical
  mechanics: a pedagogical approach. \emph{European Journal of Physics}
  41(4):045101

\bibitem[{{te Vrugt} and Wittkowski(2022)}]{teVrugtW2022}
{te Vrugt} M, Wittkowski R (2022) Perspective: New directions in dynamical
  density functionality theory. \emph{in preparation}

\bibitem[{{te Vrugt} et~al.(2020){te Vrugt}, L{\"o}wen, and
  Wittkowski}]{teVrugtLW2020}
{te Vrugt} M, L{\"o}wen H, Wittkowski R (2020) Classical dynamical density
  functional theory: from fundamentals to applications. \emph{Advances in
  Physics} 69(2):121--247

\bibitem[{{te Vrugt} et~al.(2021){te Vrugt}, Hossenfelder, and
  Wittkowski}]{teVrugtHW2021}
{te Vrugt} M, Hossenfelder S, Wittkowski R (2021) Mori-{Z}wanzig formalism for
  general relativity: a new approach to the averaging problem. \emph{Physical
  Review Letters} 127:231101

\bibitem[{{te Vrugt} et~al.(2022{\natexlab{a}}){te Vrugt},
  Frohoff-H{\"u}lsmann, Heifetz, Thiele, and Wittkowski}]{teVrugtFHHTW2022}
{te Vrugt} M, Frohoff-H{\"u}lsmann T, Heifetz E, Thiele U, Wittkowski R
  (2022{\natexlab{a}}) From a microscopic inertial active matter model to the
  {S}chr\"odinger equation. \emph{arXiv:220403018}

\bibitem[{{te Vrugt} et~al.(2022{\natexlab{b}}){te Vrugt}, Holl, Koch,
  Wittkowski, and Thiele}]{teVrugtHKWT2021}
{te Vrugt} M, Holl MP, Koch A, Wittkowski R, Thiele U (2022{\natexlab{b}})
  Derivation and analysis of a phase field crystal model for a mixture of
  active and passive particles. \emph{in preparation}

\bibitem[{Thiele et~al.(2019)Thiele, Frohoff-H{\"u}lsmann, Engelnkemper,
  Knobloch, and Archer}]{ThieleFHEKA2019}
Thiele U, Frohoff-H{\"u}lsmann T, Engelnkemper S, Knobloch E, Archer AJ (2019)
  First order phase transitions and the thermodynamic limit. \emph{New Journal
  of Physics} 21(12):123021

\bibitem[{Thompson(2016)}]{Thompson2016}
Thompson N (2016) Is naturalness natural? \emph{American Philosophical
  Quarterly} 53(4):381--395

\bibitem[{T\'{o}th(2022)}]{Toth2020}
T\'{o}th GI (2022) Emergent pseudo time-irreversibility in the classical
  many-body system of pair interacting particles. \emph{arXiv:200903089v5}

\bibitem[{Uhlenbeck and Ford(1963)}]{UhlenbeckF1963}
Uhlenbeck G, Ford G (1963) \emph{Lectures in Statistical Mechanics}. American
  Mathematical Society, Providence

\bibitem[{Verlinde(2011)}]{Verlinde2011}
Verlinde E (2011) On the origin of gravity and the laws of {N}ewton.
  \emph{Journal of High Energy Physics} 2011(4):29

\bibitem[{Vinjanampathy and Anders(2016)}]{VinjanampathyA2016}
Vinjanampathy S, Anders J (2016) Quantum thermodynamics. \emph{Contemporary
  Physics} 57(4):545--579

\bibitem[{Wallace(2010)}]{Wallace2010}
Wallace D (2010) Gravity, entropy, and cosmology: in search of clarity.
  \emph{British Journal for the Philosophy of Science} 61(3):513--540

\bibitem[{Wallace(2011)}]{Wallace2011}
Wallace D (2011) \emph{The Logic of the Past Hypothesis}. In: Loewer B,
  Winsberg E, Weslake B (eds) Time's Arrows and the Probability Structure of
  the World, Harvard (forthcoming), available at
  \texttt{\seqsplit{http://philsci-archive.pitt.edu/8894/}}

\bibitem[{Wallace(2014)}]{Wallace2014}
Wallace D (2014) Thermodynamics as control theory. \emph{Entropy}
  16(2):699--725

\bibitem[{Wallace(2015)}]{Wallace2015}
Wallace D (2015) The quantitative content of statistical mechanics.
  \emph{Studies in History and Philosophy of Modern Physics} 52:285--293

\bibitem[{Wallace(2018)}]{Wallace2018}
Wallace D (2018) The case for black hole thermodynamics part {I}:
  {P}henomenological thermodynamics. \emph{Studies in History and Philosophy of
  Modern Physics} 64:52--67

\bibitem[{Wallace(2019)}]{Wallace2019}
Wallace D (2019) The case for black hole thermodynamics part {II}:
  {S}tatistical mechanics. \emph{Studies in History and Philosophy of Modern
  Physics} 66:103--117

\bibitem[{Wallace(2021)}]{Wallace2016}
Wallace D (2021) \emph{Probability and irreversibility in modern statistical
  mechanics: classical and quantum}. In: Bedingham D, Maroney O, Timpson C
  (eds) Quantum Foundations of Statistical Mechanics, Oxford University Press,
  Oxford (forthcoming), available at arXiv:2104.11223

\bibitem[{Weisskopf(1977)}]{Weisskopf1977}
Weisskopf VF (1977) About liquids. \emph{Transactions of the New York Academy
  of Sciences} 38(1 Series II):202--218

\bibitem[{Wendt(1974)}]{Wendt1974}
Wendt RP (1974) Simplified transport theory for electrolyte solutions.
  \emph{Journal of Chemical Education} 51(10):646

\bibitem[{Winsberg(2004)}]{Winsberg2004}
Winsberg E (2004) Can conditioning on the “past hypothesis” militate
  against the reversibility objections? \emph{Philosophy of Science}
  71(4):489--504

\bibitem[{Wittkowski et~al.(2012)Wittkowski, L{\"o}wen, and
  Brand}]{WittkowskiLB2012}
Wittkowski R, L{\"o}wen H, Brand HR (2012) Extended dynamical density
  functional theory for colloidal mixtures with temperature gradients.
  \emph{Journal of Chemical Physics} 137(22):224904

\bibitem[{Wittkowski et~al.(2013)Wittkowski, L{\"o}wen, and
  Brand}]{WittkowskiLB2013}
Wittkowski R, L{\"o}wen H, Brand HR (2013) Microscopic approach to entropy
  production. \emph{Journal of Physics A: Mathematical and Theoretical}
  46(35):355003

\bibitem[{Wittkowski et~al.(2014)Wittkowski, Tiribocchi, Stenhammar, Allen,
  Marenduzzo, and Cates}]{WittkowskiTSAMC2014}
Wittkowski R, Tiribocchi A, Stenhammar J, Allen RJ, Marenduzzo D, Cates ME
  (2014) Scalar $\phi^4$ field theory for active-particle phase separation.
  \emph{Nature Communications} 5(13):4351

\bibitem[{Yoshimori(2005)}]{Yoshimori2005}
Yoshimori A (2005) Microscopic derivation of time-dependent density functional
  methods. \emph{Physical Review E} 71(3):031203

\bibitem[{Zeh(2007)}]{Zeh1989}
Zeh HD (2007) \emph{The physical basis of the direction of time (5th ed.)}.
  Springer Verlag, Berlin Heidelberg

\bibitem[{Zwanzig(1960)}]{Zwanzig1960}
Zwanzig R (1960) Ensemble method in the theory of irreversibility.
  \emph{Journal of Chemical Physics} 33(5):1338--1341

\end{thebibliography}
\end{document}